\documentclass[12pt,a4paper,openbib]{article}
\usepackage{amssymb}

\newcommand{\reseteqn}{\setcounter{equation}{0}\setcounter{subsection}{0}
  \setcounter{subsubsection}{0}}

\newcommand{\be}{\begin{equation}}
\newcommand{\ee}{\end{equation}}
\newcommand{\ba}{\begin{array}}
\newcommand{\ea}{\end{array}}
\newcommand{\bea}{\begin{eqnarray}}
\newcommand{\eea}{\end{eqnarray}}
\newcommand{\beas}{\begin{eqnarray*}}
\newcommand{\eeas}{\end{eqnarray*}}
\newcommand{\beaa}{\begin{equation}\begin{array}{lcl}}
\newcommand{\eeaa}{\end{array}\end{equation}}

\newcommand\cmp[3]   {{{\it Comm.Math.Phys.\,}{\bf #1} (#2) #3}}

\newcommand\epjc[3]  {{{\it Eur.Phys.J.\,}{\bf C#1} (#2) #3}}
\newcommand\epl[3]   {{{\it Europhys.Lett.\,}{\bf C#1} (#2) #3}}

\newcommand\jpha[3]  {{{\it J.Phys.\,}{\bf A #1} (#2) #3}}

\newcommand\grg[3]   {{{\it Gen.Rel.Grav.\,}{\bf #1} (#2) #3}}

\newcommand\jhep[3]  {{{\it J.High\,Energy\,Phys.\,}{\bf #1} (#2) #3}}
\newcommand\lmp[3]   {{{\it Lett.Math.Phys.\,}{\bf #1} (#2) #3}}

\newcommand\npb[3]   {{{\it Nucl.Phys.\,}{\bf B#1} (#2) #3}}

\newcommand\plb[3]   {{{\it Phys.Lett.\,}{\bf B#1} (#2) #3}}

\newcommand\prd[3]   {{{\it Phys.Rev.\,}{\bf D#1} (#2) #3}}

\newcommand\prl[3]   {{{\it Phys.Rev.Lett.\,}{\bf #1} (#2) #3}}

\newcommand\mpla[3]  {{{\it Mod.Phys.Lett.\,}{\bf A#1} (#2) #3}}

\newcommand\ibid[3]  {{{\it ibid.\,}{\bf #1} (#2) #3}}
\newcommand\ijmpa[3] {{{\it Int.J.Mod.Phys.\,}{\bf A#1} (#2) #3}}

\newcommand{\hepth}[1]{{\tt hep-th/#1}}

\newcommand{\condmat}[1]{{\tt cond-mat/#1}}

\newcommand{\grqc}[1]{{\tt gr-qc/#1}}

\newcommand\email[1] {{\tt#1}}

\if@twoside
\else
\fi
\def\bra{\langle}
\def\ket{\rangle}
\def\vv{\!\!\!\!\!\!\!\!\!\!\!\!}
\def\ww{\!\!\!\!\!\!}
\def\ts{\textstyle}

\def\ts{\textstyle}

\def\tfrac#1#2{{\textstyle\frac#1#2}}
\def\mb#1{\mbox{\boldmath$#1$}}

\def\nop#1{\mbox{:$#1$:}}
\def\PI{\mb{\Pi}}
\def\SU{\mb{\Sigma}}

\def\si#1#2{\sigma_{#1}^{#2}}

\begin{document}
\pagestyle{empty}

\begin{center}
  \let\footnoterule\relax
  {\LARGE\sc Operator Product Expansion in\\[1ex]
    Logarithmic Conformal Field Theory}\\[3cm]
  Mic$\hbar$ael Flohr\footnote{Research supported by EU TMR network no.\
    FMRX-CT96-0012 and the DFG String network (SPP no.\ 1096), Fl 259/2-1.}\\
  {\em Institute for Theoretical Physics, University of Hannover}\\
  {\em Appelstra\ss e 2, D-30167 Hannover, Germany}\\
  {\em E-mail: \email{flohr@itp.uni-hannover.de}}\\[1.5em]
  {\small March 20, 2002}\\[3cm]
\end{center}

\abstract{\noindent In logarithmic conformal field theory, primary fields come
  together with logarithmic partner fields on which the stress-energy
  tensor acts non-diagonally. Exploiting this fact and global conformal
  invariance of two- and three-point functions,
  operator product expansions of logarithmic operators in arbitrary rank
  logarithmic conformal field theory 
  are investigated. Since the precise relationship between logarithmic 
  operators and their primary partners is not yet sufficiently understood in
  all cases, the derivation of operator product expansion formul\ae\ is
  only possible under certain assumptions. The easiest cases are studied
  in this paper: firstly, where operator product expansions of two
  primaries only contain primary fields, secondly, where the primary
  fields are pre-logarithmic operators. Some comments on generalization towards
  more relaxed assumptions are made, in particular towards the case where
  logarithmic fields are not quasi-primary. We identify an algebraic
  structure generated by the zero modes of the fields, which proves
  useful in
  \linebreak\phantom{x}\hfill 
    determining settings in which our approach can be successfully applied.
  \hfill\phantom{x}\linebreak\phantom{x}
  }

\newpage
\pagestyle{plain}
\setcounter{page}{1}
\section{\reseteqn{\sc Introduction}}

  {\sc During the last} few years,
  so-called logarithmic conformal field
  theory (LCFT) established itself as a well-defined new animal in
  the zoo of conformal field theories in two dimensions. To our knowledge,
  logarithmic singularities in correlation functions were first noted by
  Knizhnik back in 1987 \cite{Knizhnik:1987}. 
  Although many features such as logarithmic divergences of correlators and
  indecomposable representations were observed in various places, most 
  notably in \cite{Saleur:1991,Rozansky:1992}, 
  it took six years, until the concept
  of a conformal field theory with logarithmic divergent behavior was
  introduced by Gurarie \cite{Gurarie:1993}. 
  From then one,
  there has been a considerable amount of work on analyzing the general
  structure of LCFTs, which by now has generalized almost all of the
  basic notions and tools of (rational) conformal field theories, such
  as null vectors, characters, partition functions, fusion rules,
  modular invariance etc.,
  to the logarithmic case, see for example 
  \cite{Flohr:1996a,Kausch:1995,Gaberdiel:1996a,
    RahimiTabar:1997b,Rohsiepe:1996,Ghezelbash:1997,Kogan:1998a,
    RahimiTabar:1998a,Eholzer:1997,Khorrami:1998a,Mavromatos:1998,
    MoghimiAraghi:2000a,Giribet:2001} and 
  references therein. Besides the best understood main example of the
  logarithmic $c=-2$ theory and its $c_{p,1}$ relatives, other specific 
  models were considered such as WZW models \cite{Bernard:1997,Kogan:1997,
    Nichols:2001a,Nichols:2001b,Gaberdiel:2001} and LCFTs related
  to supergroups and supersymmetry \cite{Rozansky:1992,Caux:1997,
    Khorrami:1998b,Kheirandish:2001a,Ludwig:2000,Bhaseen:2001,
    Read:2001,Kogan:2001b}.

  Also, quite a number of applications have already been pursued, and
  LCFTs have emerged in many different areas by now.
  Sometimes, longstanding puzzles in the description of certain
  theoretical models could be resolved, e.g.\ the Haldane-Rezayi state
  in the fractional quantum Hall effect \cite{Gurarie:1997,Cappelli:1998,
  Read:1999},
  multi-fractality \cite{Caux:1998b}, or two-dimensional conformal turbulence 
  \cite{Flohr:1996c,RahimiTabar:1997a,Skoulakis:1998}. 
  Other applications worth mentioning are
  gravitational dressing \cite{Bilal:1994a}, 
  polymers and abelian sandpiles \cite{Saleur:1991,Ivashkevich:1998,Cardy:1999,
    Mahieu:2001}, the (fractional) quantum Hall effect \cite{Flohr:1996b,
    Ino:1997,Kogan:2000a}, and -- perhaps most importantly -- disorder
  \cite{Caux:1996,Kogan:1996b,Maassarani:1997,Gurarie:1998,Caux:1998a,
    RahimiTabar:1998b,Gurarie:1999,Bhaseen:2000a,Bhaseen:2000b}.
  Finally, there are even applications in string theory \cite{Kogan:1996a},
  especially in 
  $D$-brane recoil \cite{Ellis:1996,Kogan:1996c,Ellis:1999,
    Mavromatos:1999a,Leontaris:1999,CampbellSmith:2000a,Lewis:2000a,
    Gravanis:2001}, 
  AdS/CFT correspondence \cite{Ghezelbash:1999,Kim:1998,Kaviani:1999,
    Kogan:1999,Myung:1999,Kogan:2000c,Sanjay:2000,MoghimiAraghi:2001b}, 
  as well as in Seiberg-Witten solutions to supersymmetric Yang-Mills theories,
  e.g.\ \cite{Flohr:1998b},
  Last, but not least, a recent focus of research on LCFTs is in its
  boundary conformal field theory aspects \cite{MoghimiAraghi:2000a,
    Kogan:2000b,Lewis:2000b,Ishimoto:2001,Kawai:2001}.

  However, the computation of correlation functions within an LCFT
  still remains difficult, and only in a few cases, four-point functions
  (or even higher-point functions) could be obtained explicitly.
  The main reason for this obstruction is that the representation theory
  of the Virasoro algebra is much more complicated in the LCFT case due
  to the fact that there exist indecomposable but non-irreducible
  representations (Jordan cells). This fact has many wide ranging
  implications. First of all, it is responsible for the appearance of
  logarithmic singularities in correlation functions. Furthermore, it
  makes it necessary to generalize almost every notion of (rational) conformal
  field theory, e.g.\ characters, highest-weight modules, null vectors etc.
  We note here that indecomposable representations need not occur with
  respect to the Virasoro algebra, but may occur with respect to (part of)
  a extended chiral symmetry algebra such as current algebras or 
  ${\cal W}$-algebras. For the sake of simplicity, we will confine 
  ourselves in this paper to the case where Jordan cells are with
  respect to the Virasoro algebra.

  In particular, what was lacking so far is a consistent generic form
  of operator product expansions (OPEs) between arbitrary rank
  logarithmic fields. Although such OPEs can be derived from
  co-product considerations in the purely representation theoretical
  framework \cite{Kausch:1995,Gaberdiel:1996a}, a direct approach trying to fix
  the generic form from global conformal covariance of the fields is
  clearly desirable. For the simple case of a rank two LCFT, where
  Jordan cells are two-dimensional, it was known since some time
  \cite{Gurarie:1993,Caux:1996} that the two-point functions of a primary
  $\Psi_{(h;0)}(z)$ and its only logarithmic partner $\Psi_{(h;1)}(z)$ are
  \bea\label{eq:examp}
    \bra\Psi_{(h;0)}(z)\Psi_{(h;0)}(w)\ket &=& 0\,,\nonumber\\
    \bra\Psi_{(h;0)}(z)\Psi_{(h;1)}(w)\ket &=& D_{(h,h;1)}(z-w)^{-2h}\,,\\
    \bra\Psi_{(h;0)}(z)\Psi_{(h;1)}(w)\ket &=& [D_{(h,h;2)} -
      2D_{(h,h;1)}\log(z-w)](z-w)^{-2h}\,.\nonumber
  \eea
  However, as we shall see, even in this simple case OPEs turn out to be
  more complicated, and one needs all possible three-point functions as well.
  First results in this direction can be found in
  \cite{RahimiTabar:1997b,Ghezelbash:1997,Khorrami:1998a,MoghimiAraghi:2001a}. 
  Here we will start to close this gap and provide the general
  structure of OPEs for fields constituting arbitrary rank Jordan cells 
  under certain assumptions.

  Let us briefly outline the basic problem: In ordinary conformal field theory,
  the generic structure of the operator product expansion is fixed up to
  structure constants which depend only on the conformal weights of the fields
  involved,
  \be
    \Phi_{h_i}(z)\Phi_{h_j}(w) = \sum_{k} C_{ij}^{\ k}\,(z-w)^{h_k}\left(
    \Phi_{h_k}(w) + \sum_{\{n\}}\beta_{ij}^{\ k,\{n\}}(z-w)^{|\{n\}|}
    \Phi_{h_k}^{(-\{n\})}(w)\right)\,.
  \ee
  Here, the fields $\Phi_h$ are primaries, and the coefficients 
  $\beta_{ij}^{\ k,\{n\}}$ of the descendant contributions
  $\Psi_{h}^{(-\{n\})} = L_{-\{n\}}\Phi_h = 
  L_{-n_1}L_{-n_2}\ldots L_{-n_l}\Phi_h$ are entirely fixed by conformal 
  covariance. The point is that the structure constants
  $C_{ij}^{\ k}$ can be easily determined if the two- and three-point 
  functions are known. In fact, these define constants 
  $D_{ij} = \bra\Psi_{h_i}(\infty)\Psi_{h_j}(0)\ket$ and 
  $C_{ijk}=\bra\Psi_{h_i}(\infty)\Psi_{h_j}(1)\Psi_{h_k}(0)\ket$ respectively,
  where $D_{ij}$ is usually a diagonal matrix, i.e.\ 
  $D_{ij}\propto\delta_{h_i,h_j}$. The two-point functions define a metric on 
  the space of fields, such that this metric $D_{ij}$ and its inverse can be 
  used to lower and raise indices (in field space) respectively. In particular,
  the OPE structure constants are simply given by
  \be\label{eq:nope}
    C_{ij}^{\ k}=C_{ijl}D^{lk}\,.
  \ee
  Now, in logarithmic conformal field theory,
  the metric induced by the two-point functions is no longer diagonal -- see
  (\ref{eq:examp}) for the simplest case, where the metric, restricted to
  fixed conformal weight $h$, is of the form
  ${0\ \ \ \ \ \ \ \ D_1\ \ \ \ \ \choose D_1\ D_2-2D_1\log(\cdot)}
  (\cdot)^{-2h}$.
  Note that, even worse, the metric cannot any longer be factorized in a 
  coordinate dependent part and a purely constant part. It is the purpose of 
  this paper to work out this metric together with all needed three-point 
  functions in order to find the correct equivalent to (\ref{eq:nope})
  in the logarithmic case.

  The paper will proceed as follows: In the next section, we will
  set up our notation and outline the different cases we will study
  in the following, progressing from the simplest setting to some more
  involved indecomposable structures. Although we restrict ourselves
  to indecomposable representations with respect to the Virasoro algebra
  alone, we will encounter a surprisingly rich set of possibilities of
  which only the easier ones can be treated with the methods presented in
  this paper.

  In the third section, we start the
  investigation in the simplest possible setting, i.e.\ where all
  logarithmic partner fields in a Jordan cell are assumed to be
  quasi-primary. This setting is as close as possible to the ordinary
  CFT case, and the generic form of one-, two-, and three-point functions
  can be computed explicitly, as long as the primary fields within the
  Jordan cells are assumed to be proper primaries. 
  We also comment on the appropriate definition of a Shapovalov form, and
  show that it is well defined under the above assumptions on the structure
  of two-point functions.
  
  In section four we derive the generic structure of OPEs in this setting.
  Although we usually work only with the holomorphic part of the LCFT,
  we briefly discuss locality in this section.
  We also compute a fully elaborated example of a rank four LCFT with only
  proper primary fields involved, demonstrating some perhaps unexpected
  features of operator products in the logarithmic case. 

  Then, in the fifth section, we consider two more general cases, 
  where the primary fields are not proper
  primaries, meaning that an OPE between two primary fields may contain
  a logarithmic field. Both examples, fermionic fields and twist fields,
  already occur in the best known LCFT, namely the $c=-2$ system, which 
  will serve as our prime example. We show that OPEs can be computed
  for these cases as well, since these primary fields are {\em not\/} part
  of Jordan cells.

  Finally, section six is devoted to the question, how the so far quite 
  restrictive assumptions can be relaxed. In particular, we investigate,
  under which circumstances logarithmic fields can be non-quasi-primary
  without affecting the results of the preceding sections. We also
  briefly address the other assumptions we made and discuss the 
  difficulties one would have to face attempting to derive the
  generic structure of OPEs and correlation functions without them.
  In particular, we demonstrate that the case of Jordan cells containing 
  primary fields which are not proper primaries cannot be solved without
  further assumptions on the structure of the CFT. Thus, we conclude
  that global conformal
  covariance fixes the generic form of correlation functions and 
  OPEs in a similar fashion as in the ordinary CFT case only when the
  indecomposable representations have a rather simple form.

  We conclude with a brief discussion of our results and possible 
  directions for future research.

\section{\reseteqn{\sc Definitions and Preliminaries}}

  {\sc To start with}, we fix some notation.
  In general, a rank $r$ Jordan cell with respect to the Virasoro algebra
  is spanned by $r$ states
  $\{|h;r-1\ket,\ldots,|h;1\ket,|h;0\ket\}$ with the property
  \begin{equation}\label{eq:L0cell}
    L_0|h;k\ket = h|h;k\ket + (1-\delta_{k,0})|h;k-1\ket\,.
  \end{equation}
  These states are defined via $\lim_{z\rightarrow 0}\Psi_{(h;k)}(z)|0\ket
  =|h;k\ket$, where $|0\ket$ denotes the SL$(2,\mathbb{C})$ invariant
  vacuum with $L_n|0\ket=0$ $\forall$ $n\geq -1$.
  The fields $\Psi_{(h;k)}(z)$ with $0<k<r$ are the so-called
  logarithmic partner fields. The field $\Psi_{(h;0)}(z)$ is a primary field
  which forms the only proper irreducible sub-representation within the 
  module of descendants of the Jordan cell.
  In the following we will denote the primary by 
  $\Phi_h(z)\equiv \Psi_{(h;0)}(z)$, if and only if it is a proper primary 
  field. We call a primary field {\em proper primary}, if its OPE with
  other proper primary fields never produces a logarithmic field on the
  right hand side. This definition is motivated from examples of LCFTs,
  where the primary fields within Jordan cells do share precisely this
  property. For instance, in the prime LCFT example, the $c=-2$ theory,
  there exists a Jordan cell of rank two for $h=0$. The primary field
  is the identity field which by definition is a proper primary.

  For completeness, we note that within a logarithmic CFT, Jordan cells
  of different rank might occur, i.e.\ $r=r(h)$ might be a function of the
  conformal weight of the corresponding (proper) primary field.
  Of course, if $r=1$, the Jordan cell
  reduces to an ordinary highest weight state, and its module of
  descendants to an ordinary Verma module. For more precise definitions
  see \cite{Rohsiepe:1996}. However, we will see later that consistency of the
  operator algebra makes it virtually impossible that Jordan cells of
  different rank occur within the same LCFT.

  We have to distinguish between the proper primary fields $\Phi_h(z)\equiv
  \Psi_{(h;0)}(z)$ in a Jordan cell and so-called pre-logarithmic
  primary fields.
  Pre-logarithmic fields are Virasoro primary fields, whose operator
  product expansions among themselves might lead to
  logarithmic fields \cite{Kogan:1998a}. Typically, pre-logarithmic fields
  turn out to be twist fields. Although no counter example is known,
  we cannot exclude that pre-logarithmic fields might occur as 
  primaries within a non-trivial Jordan cell. However, for the purposes of
  this paper, we will not assume so for the beginning. Thus, initially we
  will assume that primary fields in non-trivial Jordan cells are all proper
  primaries. Later, in section five, we will discuss the case of 
  pre-logarithmic fields.

  We also say that the field $\Psi_{(h;k)}$ has
  Jordan level $k$ in its Jordan cell, abbreviated as J-level $k$.
  Proper primary fields have J-level zero by definition. Twist fields
  do not possess a well-defined J-level. Instead, they carry a fractional charge
  $q=\ell/n$ whose denominator denotes the branching number. Logarithmic
  operators can appear in OPEs of twist fields $\chi_{h(q)}$ and $\chi_{h(q')}$,
  whenever $q+q'\in\mathbb{Z}$. 

  As discussed by F.~Rohsiepe \cite{Rohsiepe:1996}, the possible structures
  of indecomposable representations with respect to the Virasoro algebra
  are surprisingly rich. Besides the defining condition (\ref{eq:L0cell}),
  further conditions have to be employed to fix the structure. The simplest
  case is defined via the additional requirement
  \begin{equation}\label{eq:L0qp}
    L_{1}|h;k\ket = 0\,,\ \ \ \ 0\leq k<r\,.
  \end{equation}
  This condition means that all fields spanning the Jordan cell are
  quasi-primary. It will be our starting point in the following.

  The next complicated case is where logarithmic partners are not
  necessarily quasi-primary. An example is again provided by the
  simplest explicitly known LCFT, the $c=-2$ model. It possesses a
  rank two Jordan cell at $h=1$ where the logarithmic partner is not
  quasi-primary, i.e.\ $L_0|h=1;1\ket=|h=1;1\ket + |h=1;0\ket$ and
  $L_{1}|h=1;1\ket=|\xi\ket\neq0$, see \cite{Gaberdiel:1996a} for details.
  In general, this spoils attempts to make use of the global conformal
  Ward identities, since these are consequences of global conformal
  covariance or, in other words, of quasi-primarity of the fields.
  However, it seems that the condition of quasi-primary fields can
  be relaxed under certain circumstances: If the state $|\xi\ket$ generated by
  the action of $L_1$ on $|h=1;1\ket$ does not have a non-vanishing
  product with any of the states involved so far, it will not affect the
  form of the conformal Ward identities. In the given example, this is the
  case. We will postpone a more detailed discussion of this to the
  last section, but will give some definitions motivated by an
  example, which will prove useful for this later discussion.

  \subsection{{\sc An Example and Zero Mode Content}}

  In all known examples of LCFTs, the occurrence of indecomposable 
  representations can be traced back to certain pairs of conjugate zero modes.
  Let us explain this within the example of the LCFT with central
  charge $c=c_{2,1}=-2$. It may be described, see 
  for instance \cite{Gurarie:1997}, by two
  anti-commuting spin zero fields or ghost fields
  $\theta_{\alpha}$, $\alpha=\pm$, with the $SL(2,\mathbb{C})$ invariant action
  \begin{equation}
    S \propto {\rm i}\int{\rm d}^2z \varepsilon^{\alpha\beta}
    \partial\theta_{\alpha}\bar{\partial}\theta_{\beta}\,.
  \end{equation}
  In order to quantize this theory, one has to compute the fermionic
  functional integral
  \begin{equation}
    {\cal Z} = \int{\cal D}\theta_0{\cal D}\theta_1\exp(-S)\,.
  \end{equation}
  This fermionic path integral, when computed formally, vanishes due to the
  zero modes of the $\theta$ fields, which do not enter the action. To make
  it non-zero, the zero-modes have to be added by inserting the $\theta$
  fields into correlation functions, 
  \begin{equation}
    {\cal Z}' = \tfrac12\int{\cal D}\theta_0{\cal D}\theta_1
    \varepsilon^{\alpha\beta}\theta_{\alpha}\theta_{\beta}\exp(-S) = 1\,.
  \end{equation}
  As an immediate consequence, the vacuum of this theory behaves in an
  unusual way, since its norm vanishes, $\bra 0|0\ket = 0$, while the
  explicit insertion of the $\theta$ fields produces a non-zero result,
  $\frac12\bra\varepsilon^{\alpha\beta}\theta_{\alpha}(z)\theta_{\beta}(w)\ket
  = 1$. More generally, these insertions are also necessary when arbitrary
  correlation functions involving only derivatives $\partial\theta$ are
  computed, since derivatives cancel the zero modes, i.e.\ the constant
  parts of the $\theta$ fields. Therefore, we have
  \begin{equation}
    \tfrac12\bra\varepsilon^{\alpha\beta}\partial\theta_{\alpha}(z)
    \partial\theta_{\beta}(w)\ket = 0\,,\ \ \ \ {\rm but}\ \ \ \
    \tfrac14\bra\varepsilon^{\alpha\beta}\partial\theta_{\alpha}(z)
    \partial\theta_{\beta}(w)\varepsilon^{\gamma\delta}\theta_{\gamma}(0)
    \theta_{\delta}(0)\ket = \frac{-1}{(z-w)^2}\,,
  \end{equation}
  where the latter correlator is computed in analogy to the free bosonic
  field. From the viewpoint of conformal field theory, this strange
  behavior can be explained in terms of logarithmic operators which naturally
  appear in the $c=-2$ theory. As shown in \cite{Gurarie:1993}, this CFT must
  necessarily possess an operator $\tilde{\mathbb{I}}$ of scaling dimension zero  in addition to the unit operator $\mathbb{I}$, such that $[L_0,
  \tilde{\mathbb{I}}] = \mathbb{I}$ with $L_0$ (half of) the Hamiltonian.
  This property alone necessarily leads to the correlation functions
  \begin{equation}
    \bra\mathbb{I}\mathbb{I}\ket = 0\,,\ \ \ \
    \bra\mathbb{I}(z)\tilde{\mathbb{I}}(w)\ket = 1\,,\ \ \ \
    \bra\tilde{\mathbb{I}}(z)\tilde{\mathbb{I}}(w)\ket = -2\log(z-w)\,,
  \end{equation}
  which can be proved by general arguments of conformal field theory such as
  global conformal invariance and the operator product expansion.
  Furthermore, it follows that the field $\tilde{\mathbb{I}}$ can indeed
  be identified with the normal ordered product of the $\theta$ fields, i.e.
  \begin{equation}
    \tilde{\mathbb{I}}(z) = -\tfrac12\varepsilon^{\alpha\beta}
    \nop{\theta_{\alpha}\theta_{\beta}}(z)\,.
  \end{equation}

  The stress energy tensor of this CFT is given by the normal ordered
  product
  \begin{equation}\label{eq:T}
    T(z) = \tfrac12\varepsilon^{\alpha\beta}
    \nop{\partial\theta_{\alpha}\partial\theta_{\beta}}(z)\,,
  \end{equation}
  and is easily seen to fulfill the correct operator product expansion
  with itself yielding the central charge $c=-2$. To fix notation, the
  mode expansions of the $\theta$ fields read
  \begin{equation}\label{eq:theta}
    \theta_{\alpha}(z) = \xi_{\alpha} + \theta_{\alpha,0}\log(z)
    + \sum_{n\neq 0}\theta_{\alpha,n}z^{-n}\,,
  \end{equation}
  where the $\xi$'s are the crucial zero modes. The above mode expansion is
  valid in the untwisted sector (periodic boundary conditions), where
  $n\in\mathbb{Z}$. In the twisted sector (anti-periodic boundary
  conditions) $n\in\mathbb{Z}+\frac12$, and no zero modes are present.
  The anti-commutation relations read for the case $\alpha\neq\beta$ in both
  sectors
  \begin{equation}\label{eq:thth}\begin{array}{rcl}
    \{\theta_{\alpha,n},\theta_{\beta,m}\} &=& \frac{1}{n}\delta_{n+m,0}\ \ \
      {\rm for}\ \ \ n \neq 0\,,\\[0.2cm]
    \{\theta_{\alpha,0},\theta_{\beta,0}\} &=& 
    \{\xi_{\alpha},\xi_{\beta}\}\ \ = \ \ 0\,,\\[0.2cm]
    \{\xi_{\alpha},\theta_{\beta,0}\} &=& 1\,,
  \end{array}
  \end{equation}
  with all other anti-commutators vanishing. Note that the $\xi$-modes
  become the creation operators for logarithmic states. Indeed, the
  highest weight conditions of the standard $SL(2,\mathbb{C})$ invariant
  vacuum are
  \begin{equation}\label{eq:thhw}
    \theta_{\alpha,n}|0\ket = 0\ \ \forall\ n\geq 0\,,
  \end{equation}
  such that
  \begin{equation}
    |\tilde 0\ket = \tilde{\mathbb{I}}(0)|0\ket =
    -\tfrac12\varepsilon^{\alpha\beta}\xi_{\alpha}\xi_{\beta}|0\ket\,.
  \end{equation}
  It is instructive to conclude our example with the little exercise
  to compute $L_0|\tilde 0\ket$ by using the mode expansion
  $L_n = \frac12\sum_m\varepsilon^{\alpha\beta}
  \nop{\theta_{\alpha,n-m}\theta_{\beta,m}}$ as follows:
  \begin{equation}\begin{array}{rcl}
    L_0|\tilde 0\ket &=&
      -\frac14\varepsilon^{\alpha\beta}\varepsilon^{\gamma\delta}
      \sum_m\nop{(\delta_{m,0}-m^2)\theta_{\alpha,-m}\theta_{\beta,m}}\,
      \xi_{\gamma}\xi_{\delta}|0\ket\\[0.3cm]
    &=& +\frac14\varepsilon^{\alpha\beta}\varepsilon^{\gamma\delta}\left(
      \xi_{\gamma}\xi_{\delta}\sum_{m>0}m^2(\theta_{\alpha,-m}\theta_{\beta,m}
      - \theta_{\beta,-m}\theta_{\alpha,m})
      - \theta_{\alpha,0}\theta_{\beta,0}\xi_{\gamma}\xi_{\delta}\right)
      |0\ket\\[0.3cm]
    &=& -\frac14\varepsilon^{\alpha\beta}\varepsilon^{\gamma\delta}
      \theta_{\alpha,0}\theta_{\beta,0}\xi_{\gamma}\xi_{\delta}|0\ket
    \ \ =\ \ \frac12\varepsilon^{\alpha\beta}\varepsilon_{\alpha\beta}|0\ket
    \ \ =\ \ |0\ket\,,
  \end{array}
  \end{equation}
  where the third equality follows from the highest weight condition
  (\ref{eq:thhw}), and
  otherwise the anti-commutation relations (\ref{eq:thth}) were used.
  This clearly demonstrates that the states $|\tilde 0\ket$ and
  $|0\ket$ form an indecomposable Jordan cell with respect to the
  Virasoro algebra. The action of other Virasoro modes can be
  computed in the same fashion.
  
  What we learn from this example is the crucial role of 
  conjugate pairs of zero modes, i.e.\ pairs $\xi_{\alpha}$, $\theta_{\beta,0}$.
  It turns out that logarithmic fields are precisely those fields,
  whose mode expansion contains $\varepsilon^{\alpha\beta}\xi_{\alpha}
  \xi_{\beta}$. Therefore, it makes sense to talk of the ``logarithmicity''
  of a field, more precisely of its $\xi$ zero mode content.
  It is clear that there might be more pairs of conjugate modes in more
  general LCFTs (e.g.\ in
  higher spin ghost system) and hence a field might possess a higher zero 
  mode content. The phrase zero mode is ambiguous: We do not mean
  the zero-th mode in the mode expansion of a field, but pairs of conjugate
  modes $a_n$, $c_{-n}$ such that $a_n$ annihilates to both sides, i.e.\
  $\bra 0|a_n=a_n|0\ket=0$, and
  $c_{-n}$ is a creator to both sides. In the example, $\theta_{\alpha,0}$
  are the annihilators and $\xi_{\beta}$ the creators. The zero mode
  content of a field counts the number of creation operator zero modes.
  If the modes are anti-commuting, we will also talk of an even or odd zero 
  mode content of fields. Fields of even zero mode content are called
  bosonic, fields of odd zero mode content fermionic, respectively.
  It is important to note that a correlation 
  function can only be non-vanishing, if its total zero mode content
  is large enough to kill all annihilator zero modes. In our example,
  any non-vanishing correlator must contain $\varepsilon^{\alpha\beta}
  \xi_{\alpha}\xi_{\beta}$. Moreover, our example is of fermionic nature,
  such that the zero mode content must always be even. 

  To fix notation,
  we will denote the number of creator zero modes of a field $\Psi$ by
  $Z_0(\Psi)$. If the modes are fermionic, we further introduce
  $Z_-(\Psi)$ and $Z_+(\Psi)$ which count the modes with respect to 
  the spin doublet label $\alpha=\pm$ respectively. Of course, 
  $Z_0(\Psi) = Z_-(\Psi)+Z_+(\Psi)$ in this case. We stress that
  correlation functions $\bra\Psi_1(z_1)\ldots\Psi_n(z_n)\ket$
  of fields $\Psi_i(z_i)$ with $Z_0(\Psi_i)=0$ for all $i=1,\ldots,n$ 
  must vanish. Moreover, if the minimal zero mode content
  of a given LCFT is $N$, all correlators vanish, whenever
  $\sum_iZ_0(\Psi_i)< N$. Thus, such fields behave almost as null
  fields. They are not entirely null, since we may insert those fields
  in a correlator with already sufficiently high total zero mode content 
  without necessarily forcing it to become zero. In our example, 
  $N = N_-+N_+ = 1+1 = 2$. The rank of Jordan cells of a LCFT generated
  by anti-commuting fields is $N/2+1$, otherwise it should be $N+1$, as
  in the case of LCFTs from puncture operators in Liouville theories
  \cite{Ellis:1996,Kogan:1996b,Kogan:1998a}.

\subsection{{\sc Zero Mode Content and J-Level}}

  We will discuss in section six that the zero mode content allows 
  to put bounds on the J-levels of fields appearing on the right
  hand side of OPEs. However, it is not clear whether all LCFTs admit
  to assign a zero mode content in the above defined sense to all of 
  its fields. Thus, the basic problem one is faced with is that there
  is no a priori rule that restricts the J-level in an OPE of the
  form
  \be\label{ope}
  \Psi_{(h_1;k_1)}(z)\Psi_{(h_2;k_2)}(w) = \sum_h\sum_{k=0}^{r(h)-1}
  C_{(h_1;k_1)(h_2;k_2)}^{(h;k)}(z-w)^{h-h_1-h_2}f_{k_1k_2}^k(z-w)
  \Psi_{(h;k)}(w)\,,
  \ee
  where the functions $f_{k_1k_2}^k(x)$ collect possible logarithmic terms.
  What one would definitely wish for would be something like a gradation
  such that the J-level of the right hand side of (\ref{ope}) is bounded as
  \be\label{jlevel}
    k \leq k_1+k_2\,.
  \ee 
  Unfortunately, we know that this is not always true. Pre-logarithmic
  fields, for instance, are true primary fields which produce a
  logarithmic field in their OPE which means that $1=k>k_1+k_2=0+0$ in
  contradiction to the above bound.

  Now, if the LCFT under consideration admits to assign a well-defined
  zero mode content to each of its fields, we can trace back the
  origin of logarithms and of Jordan cells to precisely that zero mode
  content, as indicated in the preceding subsection. Since the Jordan
  cell structure is in this case generated by the existence of certain
  creator zero modes, it is clear that the OPE of two fields
  can never produce fields on the right hand side whose zero mode
  content exceeds the initial one. This follows by the simple fact that
  the OPE can be computed on the level of modes explicitly by contraction.
  Thus, it follows that the existence of a well-defined zero mode content
  puts a natural bound on the maximal zero mode content of OPEs. Therefore,
  instead of the above inequality (\ref{jlevel}), 
  we then have for the right hand side of the OPE (\ref{ope})
  \be\label{zmc}
    Z_0(\Psi_{(h;k)}) \leq Z_0(\Psi_{(h_1;k_1)}) + Z_0(\Psi_{(h_2;k_2)})\,.
  \ee
  Since, on the other hand, the zero mode content determines the Jordan
  cell structure via the off-diagonal action of the generators of the
  chiral symmetry algebra, the inequality (\ref{zmc}) implies the 
  desired inequality (\ref{jlevel}).

  It is clear that we are now faced with a dilemma. If we don't know
  a priori which fields at which J-levels may contribute to the
  right hand side of OPEs, we do not have a way to compute the generic
  form of such OPEs, since we do not have enough information to fix
  its structure. Thus, we are forced to make even more assumptions on
  the structure of LCFTs. We will therefore assume throughout the
  whole paper that a condition of the form $(\ref{jlevel})$ holds, 
  as long as fields from Jordan cells are concerned. We will justify 
  this assumption as reasonable in section six for the case where
  the LCFT admits a description in terms of a zero mode content.
  However, we don't know whether all LCFTs can be described in this way. 

  As we will briefly discuss at the end of section six, there are
  indications that LCFTs may not be consistent, if no such condition
  restricting the maximal J-level of OPEs exists. In general, we don't
  have any means to compute the structure of arbitrary $n$-point 
  correlation functions in this case, but it is still possible to look at 
  the two-point functions. Our results indicate that these may indeed be
  inconsistent if, for example, primary members of Jordan cells can
  produce logarithmic partners in their OPE.

\section{\reseteqn{\sc SL$(2,\mathbb{C})$ Covariance}}

  {\sc In ordinary CFT}, two- and three-point functions are determined up to
  constants which determine the operator algebra and must be fixed by
  the associativity of the operator product expansion. Moreover,
  one-point functions are trivial, i.e.\ $\bra\Phi_h(z)\ket = \delta_{h,0}$,
  although Zamolodchikov pointed out a long time ago, that in non-unitary
  CFTs, non-vanishing one-point functions might be possible.
  For the beginning, we consider only the chiral half of the
  theory, but keep in mind that LCFTs are known not to factorize entirely
  into chiral and anti-chiral halfs.

  In order to find the generic structure of two- and three-point functions
  in logarithmic CFTs, we must consider different cases. We start with
  the simplest setting, as outlined in the preceeding section. 
  Throughout this section we will therefore assume the
  following: We consider correlation functions of fields 
  $\Psi_{(h_i;k_i)}(z_i)$ from Jordan cells where $\Psi_{(h_i;0)}(z_i)$ are
  proper primary fields. This assumption guarantees that the operator
  product expansion of two of the primaries will contain only primary
  fields (and their descendants). Thus, the primaries behave exactly as
  in an ordinary CFT. Furthermore, we assume that all the logarithmic fields
  in the Jordan cells are quasi-primary, i.e.\ that $L_1\Psi_{(h_i;k_i)}(0)|0
  \ket=0$ for all $k_i=0,\ldots,r(h_i)-1$. Such Jordan cells will be called
  proper Jordan cells in the following. We remark that in
  any sensible LCFT there is at least one (possibly trivial) Jordan cell
  which satisfies these assumptions, namely the $h=0$ Jordan cell where
  the primary is the identity field. The identity should exist in any
  sensible CFT, since it is the unique field associated with the
  $SL(2,\mathbb{C})$-invariant vacuum. 

  Under these assumptions, as shown in the latter two references in 
  \cite{Flohr:1996a}, the action of the Virasoro modes receives an additional
  non-diagonal term, namely
  \begin{eqnarray}\label{eq:vir}
    & & L_n \bra\Psi_{(h_1;k_1)}(z_1)\ldots\Psi_{(h_n;k_n)}(z_n)\ket =
    \nonumber\\
    & & \sum_iz_i^n\left[z_i\partial_i + (n+1)(h_i+\hat{\delta}_{h_i})\right]
      \bra\Psi_{(h_1;k_1)}(z_1)\ldots\Psi_{(h_n;k_n)}(z_n)\ket
  \end{eqnarray}
  where $n\in\mathbb{Z}$ and the off-diagonal action is
  $\hat{\delta}_{h_i}\Psi_{(h_j;k_j)}(z) = \delta_{ij}\Psi_{(h_j;k_j-1)}(z)$ for
  $k_j>0$ and $\hat{\delta}_{h_i}\Psi_{(h_j;0)}(z) = 0$. This little extension
  has tremendous consequences. As we are going to show, even the
  simplest quantities, namely the one-point functions, are severely
  modified in their behavior. To start with, we recall that only infinitesimal
  conformal transformations in the algebra $\mathfrak{sl}(2,\mathbb{C})$ can be
  integrated to global conformal transformation on the Riemann sphere.
  Thus, only the generators $L_{-1}$, $L_0$, and $L_1$ of the M\"obius
  group admit globally valid conservation laws, which usually are expressed
  in terms of the so-called conformal Ward identities
  \begin{equation}\label{eq:ward} 0 = \left\{\begin{array}{rcl}
    L_{-1} G(z_1,\ldots z_n) & = & \sum_i\partial_i G(z_1,\ldots z_n)
      \,,\\[0.2cm]
    L_0 G(z_1,\ldots z_n) & = & \sum_i(z_i\partial_i + h_i +
      \hat{\delta}_{h_i})G(z_1,\ldots z_n)\,,\\[0.2cm]
    L_1 G(z_1,\ldots z_n) & = & \sum_i(z_i^2\partial_i
      + 2z_i[h_i + \hat{\delta}_{h_i}]) G(z_1,\ldots z_n)\,,
    \end{array}\right.
  \end{equation}
  where $G(z_1,\ldots z_n)$ denotes an arbitrary $n$-point function
  $\bra\Psi_{(h_1;k_1)}(z_1)\ldots\Psi_{(h_n;k_n)}(z_n)\ket$ of primary
  fields and/or their logarithmic partner fields. Here, we already have
  written down the Ward identities in the form valid for proper
  Jordan cells in logarithmic conformal field theories. We will see in
  section six, that the assumption of quasi-primary logarithmic fields
  can be relaxed under certain circumstances.

  \subsection{{\sc One-point Functions}}

  Let us now apply the Ward identities (\ref{eq:ward}) to an arbitrary
  one-point function $G(z) = \bra\Psi_{(h;k)}(z)\ket$ of a field in a
  rank $r$ Jordan cell. The identity for
  $L_{-1}$ states translational invariance such that $G(z)=E_{(h;k)}$ must
  be a constant independent of the position $z$. But the identity for $L_0$,
  stating scaling and rotational invariance, leads to the condition
  \begin{equation}\label{eq:1pt}
    hE_{(h;k)} + (1-\delta_{k,0})E_{(h;k-1)} = 0\,.
  \end{equation}
  In case of the one-point
  functions, special conformal transformations do not yield an additional
  constraint. However, the above condition immediately results in the
  recursive relation, $E_{(h;r-1-l)} =
  (-h)^lE_{(h;r-1)}$, such that, if $E_{(h;r-1)}$ is non-zero, automatically
  all other one-point functions in this Jordan cell also do not vanish,
  as long as $h\neq 0$. For $h=0$, the only non-vanishing one-point
  function is the one of highest possible J-level, i.e.\ $E_{(h;r-1)}\neq 0$,
  $E_{(h;k)} = 0$ for $0\leq k < r-1$. Note that $E_{(h=0;r-1)}$ must be
  non-zero. Otherwise, the whole Jordan module to fields of scaling dimension
  zero could be removed from the theory, since it were orthogonal to all
  other states. Then, the remaining CFT would not have a vacuum state.
  To be specific, we from now on normalize $E_{(0;r-1)}=1$.

  We can learn one more thing from the one-point functions: If a field
  $\Psi_{(h;k)}(z)$, $0<k<r$, were not quasi-primary, special conformal
  transformations yield a non-zero result as long as the expectation
  value of the non-quasi-primary contribution is non-zero. Namely,
  $$
    L_1\bra\Psi_{(h;k)}(z)\ket = (-h)^{r-1-k}L_1E_{(h;r-1)} = 
    \bra\Psi'_{(h-1)}(z)\ket
  $$ 
  for a suitable (not necessarily quasi-primary) field $\Psi'(z)$ with
  conformal weight $h-1$ (which is not necessarily part of a Jordan cell,
  which is why we omit the J-level). We will see in section six that one 
  property of the off-diagonal action of the Virasoro modes is to reduce 
  the zero mode content, i.e.\ $Z_0(\Psi')<Z_0(\Psi)$. Hence, instead of
  quasi-primarity, a weaker condition will always hold, namely
  $L_1^{n(h,k)}\Psi_{(h;k)}(0)|0\ket = 0$ for a certain
  $n(h,k)$ depending on the conformal weight $h$ and the J-level $k$.
  Moreover, the reduced zero mode content may already force the
  vacuum expectation value to vanish, i.e.\ the situation may arise
  that $L_1\Psi_{(h;k)}(0)|0\ket\neq 0$, but $L_1\bra\Psi_{(h;k)}(z)\ket=0$.
  This is precisely the case for the example given in section II.1, namely
  the Jordan cell at $h=1$. In particular, $n(h=1,1)=2$ in the $c=-2$ case, 
  i.e.\ $L_1^2|h=1;1\ket=L_1|\xi\ket = 0$ but $\bra\xi\ket = 0$. 

  \subsection{{\sc Two-point Functions}}

  The next step is to consider two-point functions $G=\bra
  \Psi_{(h_1;k_1)}(z_1)\Psi_{(h_2;k_2)}(z_2)\ket$ of two fields belonging to
  Jordan cells of ranks $r_1,r_2$ respectively.
  Translational invariance tells us
  that $G=G(z_{12})$ is a function of the distance only. Scaling invariance
  then leads to the ordinary first order differential equation
  \begin{equation}
    \left(z_{12}\partial_{z_{12}} + h_1+h_2\right)G(z_{12})
    + \bra\Psi_{(h_1;k_1-1)}(z_1)\Psi_{(h_2;k_2)}(z_2)\ket
    + \bra\Psi_{(h_1;k_1)}(z_1)\Psi_{(h_2;k_2-1)}(z_2)\ket\,.
  \end{equation}
  The generic solution to this inhomogeneous equation is already
  surprisingly complicated. Let us introduce some nomenclature to denote
  where in a correlator logarithmic partners of a primary are inserted
  by writing
  \begin{equation}
    \bra\Psi_{(h_1;k_1)}(z_1)\Psi_{(h_2;k_2)}(z_2)\ldots
    \Psi_{(h_n;k_n)}(z_n)\ket = G_{k_1,k_2,\ldots k_n}(z_1,z_2,\ldots z_n)\,.
  \end{equation}
  The above equation then becomes
  $(z_{12}\partial_{z_{12}}+h_1+h_2)G_{k_1,k_2}(z_{12}) =
  - G_{k_1-1,k_2}(z_{12}) - G_{k_1,k_2-1}(z_{12})$ with solution
  \begin{equation}
    G_{k_1,k_2}(z_{12})=(z_{12})^{-h_1-h_2}\left(D_{(h_1;k_1)(h_2;k_2)}
    - \int^{z_{12}}\!\frac{{\rm d}\zeta}{\zeta^{1-h_1-h_2}}[
    G_{k_1-1,k_2}(\zeta)+G_{k_1,k_2-1}(\zeta)]\right)\,.
  \end{equation}
  An explicit solution can be found in a hierarchical way, starting with
  the two-point function of proper primary fields, $G_{0,0}(z_1,z_2)$.
  The conformal Ward identities then reduce to the common CFT case with
  the well-known solution
  \begin{equation}
    \bra\Phi_{h_1}(z_1)\Phi_{h_2}(z_2)\ket = D_{(h_1;0)(h_2;0)}
    \delta_{h_1,h_2}(z_1-z_2)^{-h_1-h_2}\,.
  \end{equation}
  However, to be consistent with insertion of an OPE, the constant
  must satisfy $D_{(h;0)(h;0)} = C_{(h;0)(h;0)}^{(0;0)}E_{(0;0)} = 0$,
  due to our results on the one-point functions. Hence, $G_{0,0}(z_1,z_2)=0$
  and, moreover, $E_{(h;0)} = 0$ for $h\neq 0$ since the form of the
  two-point function does not admit contributions from other one-point
  functions. We conclude that the only non-vanishing one-point function
  of fields in Jordan cells is $\bra\Psi_{(0;r(0)-1)}\ket$.

  The reader should note that in the above reasoning we crucially
  made use of the assumption that the primary fields are proper, i.e.\ that
  their OPE does not contain any logarithmic fields. We will come back
  to this point later. However, if the identity of the CFT belongs to
  a Jordan cell, then the assumption that the primary be proper is
  automatically satisfied. Moreover, as already stressed earlier, 
  it is assumed throughout this section that all logarithmic partner
  fields are quasi-primary.

  We can go on and consider $G_{1,0}(z_1,z_2)$ next. The Ward identities
  now yield an additional term proportional to $G_{0,0}$, which luckily
  vanishes as just shown. Therefore, we can conclude that $G_{1,0}$ is
  non-zero, if and
  only if $E_{(0;1)}$ is non-zero, i.e.\ if and only if $r-1 = 1$.
  Going on in this manner, we finally arrive at the general statement
  \begin{equation}
    \bra\Psi_{(h_1;k)}(z_1)\Psi_{(h_2;0)}(z_2)\ket =
    \bra\Psi_{(h_1;0)}(z_1)\Psi_{(h_2;k)}(z_2)\ket =
    \delta_{h_1,h_2}\delta_{k,r-1}D_{(h_1,h_1;r-1)}\cdot(z_{12})^{-2h_1}\,,
  \end{equation}
  which does not depend on which of the two fields is the field of
  maximal J-level. It is more complicated to compute two-point functions
  where both fields have J-level larger zero, except when the Jordan rank
  is $r=2$. Then the only other possibility is $G_{1,1}(z_1,z_2)$, where
  the Ward identities yield contributions proportional to $G_{1,0}=G_{0,1}$
  with solution $\bra\Psi_{(h;1)}(z_1)\Psi_{(h;1)}(z_2)\ket =
  (z_{12})^{-2h}[D_{(h,h;2)} - 2D_{(h,h;1)}\log(z_{12})]$.

  When generalizing to arbitrary rank Jordan cells,
  the following picture emerges for the two-point functions:
  The structure constants depend only on the total J-level, i.e.\
  $D_{(h;k)(h;l)}=D_{(h;k')(h;l')}\equiv D_{(h,h;k+l)}$ for $k+l=k'+l'$,
  and they vanish, if the
  total J-level is less than the rank of the vacuum representation, i.e.\
  $D_{(h;k)(h;l)}=0$ for $k+l+1<r(h=0)$. Another consequence is that the
  only non-vanishing one-point function of type $E_{(h;k)}$ is
  $E_{(h=0;r(h=0)-1)}$. This, in turn, implies that
  a logarithmic CFT is only consistently possible, if
  the vacuum representation is a Jordan cell representation of maximal
  rank $r(h=0)\geq r(h)\ \forall h\neq 0$.
  We then say that the LCFT has rank $r$.
  Putting things together, the complete solution for the
  two-point function must have the form
  \begin{equation}\label{eq:2pt}
    \bra\Psi_{(h_1;k_1)}(z_1)\Psi_{(h_2;k_2)}(z_2)\ket =
    \delta_{h_1,h_2}\left(\sum_{\ell=0}^{k_1+k_2}D_{(h_1,h_2;k_1+k_2-\ell)}
    \frac{(-2)^{\ell}}{\ell!}\log^{\ell}(z_{12})\right)(z_{12})^{-h_1-h_2}\,,
  \end{equation}
  where we have indicated the implicit condition $h_1=h_2$ and where for
  a rank $r$ LCFT all constants $D_{(h,h;k)}=0$ for $k < r-1$. 
  This result was first obtained in \cite{RahimiTabar:1997b}. In this way,
  the two-point functions define for each possible conformal weight $h$
  matrices $G^{(2)}_{k_1,k_2}$ of size $r(h)\times r(h)$. However, these
  matrices depend only on $2r(h)-r$ yet undetermined constants
  $D_{(h,h;k)}$, $r-1\leq k\leq 2r(h)-2$. Moreover, all entries above the
  anti-diagonal are zero.
  This last property, i.e.\ that $D_{(h,h;k)}=0$ for $k < r-1$,
  is due to the one-point functions since
  \begin{eqnarray}
    D_{(h,h;k)} &=&
    \frac{1}{2r(h)-k-1}
    \sum_{{0\leq\ell_1,\ell_2\leq r(h)-1\atop\ell_1+\ell_2=k}}
    C_{(h;\ell_1)(h;\ell_2)}^{(0;r-1)}E_{(0;r-1)}\nonumber\\
    &=& \left\{
      \begin{array}{lcl}
        C_{(h,h,0;k+r-1)} & &
          {\rm for}\ \ r-1 \leq k < 2r-1\,,\\
        0 & & {\rm else}\,.
      \end{array}\right.
  \end{eqnarray}
  Note that the three-point structure constants do, in effect, only
  depend on the total J-level, as we have tried to indicate in our notation.
  The special form of the two-point structure constant matrices ensures that
  they are always invertible.

  It is often very convenient to work with states instead of the fields
  directly, in particular when purely algebraic properties such as null
  states are considered. As usual, we have an isomorphism between the
  space of fields and the space of states furnished by the map
  $|h;k\ket = \Psi_{(h;k)}(0)|0\ket$. Although one does not necessarily
  have a scalar product on the space of states, one can introduce a
  pairing, the Shapovalov form, between states and linear functionals.
  Identifying the out-states with (a subset of) the linear functionals
  equips the space of states with a Hilbert space like structure.
  As in ordinary conformal field theory, we have
  $\bra h;k| = (|h;k\ket)^{\dagger} = \lim_{z\rightarrow 0}\bra 0|
  \Psi_{(h;k)}(1/z)$. Using now that logarithmic fields transform under
  conformal mappings $z\mapsto f(z)$ as
  \begin{eqnarray*}
    \Psi_{(h;k)}(z) &=& \sum_{l=0}^k\frac{1}{l!}\frac{\partial^l}{\partial
      h^l}\left(\frac{\partial f(z)}{\partial z}\right)^h\Psi_{(h;k-l)}(f(z))\\
                    &=& \sum_{l=0}^k\frac{1}{l!}\log^l\left|\frac{\partial 
      f(z)}{\partial z}\right|\left(\frac{\partial f(z)}{\partial z}\right)^h
      \Psi_{(h;k-l)}(f(z))\,,
  \end{eqnarray*}
  the out-state can be re-expressed in a form which allows us to apply
  (\ref{eq:2pt}) to evaluate the Shapovalov form. In ordinary conformal
  field theory, we simply get $\bra h| = \lim_{z\rightarrow\infty}\bra 0|
  z^{2h}\Psi_h(z)$ such that $\bra h|h'\ket = \delta_{h,h'}$ up to
  normalization. Interestingly, the transformation behavior of
  logarithmic fields yields a very similar result, canceling all
  logarithmic divergences. Thus, we obtain for the Shapovalov form
  $$
    \bra h;k|h';k'\ket = \delta_{h;h'}D_{(h,h';k+k')}\,, 
  $$
  which is a lower triangular matrix. To demonstrate this, we
  consider the example of a rank two LCFT. Then we clearly have
  $\bra h;0|h;0\ket = 0$, $\bra h;1|h;0\ket=\bra h;0|h;1\ket = D_{(h,h;1)}$
  and with
  $$
    \lim_{z\rightarrow 0}\bra 0|\Psi_{(h;1)}(1/z)
    \Psi_{(h;1)}(0)¦\ket = \lim_{z\rightarrow\infty}\bra 0|
    z^{2h}\left[\Psi_{(h;1)}(z) + 2\log(z)\Psi_{(h;0)}(z)\right]
    \Psi_{(h;1)}(0)|0\ket
  $$
  the desired result $\bra h;1|h;1\ket = D_{(h,h;2)}$. Hence, the
  Shapovalov form is well defined and non-degenerate for the logarithmic
  case much in the same way as it can be defined for ordinary CFTs. Note that
  the definition of the Shapovalov form does not depend on whether the CFT
  is unitary or not. 

  For completeness, we mention that the Shapovalov form is not uniquely
  defined in
  LCFTs, because the basis $\{|h;k\ket : k=0,\ldots r(h)-1\}$ of states
  is not unique. The reason is that we always have the freedom to
  redefine the logarithmic partner fields, or their states respectively,
  as
  $$
    \Psi'_{(h;k)}(z) = \Psi_{(h;k)}(z) + \sum_{i=1}^k\lambda_i\Psi_{(h;k-i)}(z)
  $$
  with arbitrary constants $\lambda_i$. At this state, there are no further
  restrictions from the structure of the LCFT which could fix a basis within
  the Jordan cells. Only the proper primary field, or the proper highest-weight
  state respectively, is uniquely defined up to normalization.

  \subsection{{\sc Three-point Functions}}

  The three-point functions can be fixed along the same lines, although
  the procedure is now more complicated. For each triplet $h_1,h_2,h_3$ of
  conformal weights, we find a set of $r(h_1)\times r(h_2)\times r(h_3)$
  functions $G_{k_1,k_2,k_3}(z_1,z_2,z_3)$. From now on we will restrict
  ourselves to the case where $r(h)=r$ for all Jordan cells in the LCFT.
  We will see shortly that otherwise no consistent definition of OPEs
  seems possible. With this restriction, we can collect the
  set of three-point functions into $r$ matrices, each of size $r\times r$,
  namely the matrices $(G^{(3)}_{k_1})^{}_{k_2,k_3}$.

  A closed formula of the type as given above for the two-point function
  is extremely lengthy. However, the three-point functions can all be
  given in the form:
  \begin{eqnarray}\label{eq:3pt}
    & & \vv\bra\Psi_{(h_1;k_1)}\Psi_{(h_2;k_2)}\Psi_{(h_3;k_3)}\ket
    = \sum_{k=r-1}^{k_1+k_2+k_3}C_{(h_1,h_2,h_3;k)}
    \sum_{j_1=0}^{k_1}\sum_{j_2=0}^{k_2}\sum_{j_3=0}^{k_3}
    \delta_{j_1+j_2+j_3,k_1+k_2+k_3-k}\phantom{mmmn} \nonumber\\
    & & \phantom{mmmm}\times\
    \frac{1}{j_1!j_2!j_3!}(\partial_{h_1})^{j_1}(\partial_{h_2})^{j_2}
    (\partial_{h_3})^{j_3}\left(z_{12}^{h_3-h_1-h_2}z_{13}^{h_2-h_1-h_3}
    z_{23}^{h_1-h_2-h_3}\right)\,.
  \end{eqnarray}
  The corresponding formula for the two-point function can be rewritten
  in the same manner involving derivatives with respect to the conformal
  weight,
  \begin{eqnarray}
    & & \bra\Psi_{(h_1;k_1)}\Psi_{(h_2;k_2)}\ket
    = \sum_{k=r-1}^{k_1+k_2}\delta_{h_1,h_2}D_{(h_1,h_2;k_1+k_2-k)}
    \frac{1}{k!}(\partial_{h_2})^{k}(z_{12})^{-2h_2}\,,
  \end{eqnarray}
  which evaluates to exactly the form given in (\ref{eq:2pt}).
  Note that again the yet free structure constants depend only on the
  total J-level. This agrees with what one might expect from the total
  symmetry of the three-point structure constants under permutations.
  Differentiation with respect to the conformal weights reproduces
  precisely the logarithmic contributions to satisfy the inhomogeneous
  Ward identities.

  These expressions can be made even more suggestive, if one treats the
  structure constants as (analytic) functions of the conformal weights
  \cite{RahimiTabar:1997b}.
  This is actually true in the case of minimal models, where all structure
  constants can be given explicitly as functions of the charges within
  a free field representations, and hence in terms of the conformal weights.
  Putting simply
  \begin{equation}
     C_{(h_1,h_2,h_3;k+r-1)} = {\ts\frac{1}{k!}}\sum_{i_1+i_2+i_3=k-(r-1)}
     (\partial_{h_1})^{i_1}(\partial_{h_2})^{i_2}(\partial_{h_2})^{i_2}
     C_{h_1,h_2,h_3}\,,
  \end{equation}
  allows to rewrite (\ref{eq:3pt}) entirely in terms of derivatives with
  respect to the conformal weights. Here, $C_{(h_1,h_2,h_3;r-1)}$
  is then the pure, not differentiated, structure constant.

\section{\reseteqn{\sc Operator Product Expansions}}

  {\sc With the complete} set of two- and three-point functions at hand, we can
  now proceed to determine the operator product expansions in their generic
  form. To do this, we first consider the asymptotic limit
  $\lim_{z_1\rightarrow z_2}G^{(3)}_{k_1,k_2,k_3}(z_1,z_2,z_3)$ and define
  the matrices $(G^{(3)}_{k_1})^{}_{k_2,k_3}$ in this limit. This essentially
  amounts to replacing $z_{13}$ by $z_{23}$. Next, we take the two-point
  functions $G^{(2)}_{k_1,k_2}(z_2,z_3)$, collect them into a matrix
  $(G^{(2)})_{k_1,k_2}$ and invert the latter to obtain
  $(G^{(2)})^{\ell_1,\ell_2}$. Finally, the matrix product
  \begin{equation}\label{eq:O=CD}
    C_{(h_1,h_2;k_1+k_2)}^{(h_3;k_3)} = (G^{(3)}_{k_1})^{}_{k_2,k}
    (G^{(2)})^{k,k_3}
  \end{equation}
  yields matrices $(C_{(h_1;k_1),h_2}^{\ h_3})_{k_2}^{k_3}$ encoding all the
  OPEs of the field $\Psi_{(h_1;k_1)}(z)$ with fields of arbitrary J-level.
  An immediate consequence of (\ref{eq:O=CD}) is now that associativity of
  the operator algebra can only hold if the rank of all Jordan cells is
  equal. Indeed, assuming the contrary, the matrices $(G^{(3)}_{k})_{lm}$
  were not always square matrices, and the rank of the matrices $(G^{(2)})^{kl}$
  would depend on the conformal weight. It is now easy to see that
  the associativity conditions such as crossing symmetry
  \begin{equation}
    C_{(h_i,h_j;k_i+k_j)}^{(h;k)}C_{(h,h_l,h_m;k+k_l+k_m)}^{\vphantom{(}} =
    C_{(h_i,h_l;k_i+k_l)}^{(h';k')}C_{(h',h_j,h_m;k'+k_j+k_m)}^{\vphantom{(}}
  \end{equation}
  cannot any longer hold, since the matrices on both sides of the equation
  were not always of equal rank. In effect, associativity can only be kept
  if the ranks of the Jordan cells appearing implicitly on both sides
  of the equation can consistently be restricted to the minimal rank of
  the product matrices. This minimal rank will automatically define the
  maximal rank of the LCFT under consideration.
  This justifies our earlier restriction.

  To see, how this formula works, we will give a more explicit version of
  (\ref{eq:O=CD}). Let us denote the complete set of two--point functions
  as $\bra\ell,k\ket = G^{(2)}_{\ell,k}(z_2,z_3) = \bra
  \Psi_{(h;\ell)}(z_2)\Psi_{(h;k)}(z_3)\ket$ and correspondingly the 
  three-point functions as $\bra\ell,k_1,k_2\ket =
  \lim_{z_1\rightarrow z_2}G^{(3)}_{k_1,k_2,\ell}(z_1,z_2,z_3) =
  \lim_{z_1\rightarrow z_2}\bra\Psi_{(h_1;k_1)}(z_1)\Psi_{(h_2;k_2)}(z_2)
  \Psi_{(h;\ell)}(z_3)\ket$, all essentially
  given by formulae (\ref{eq:2pt}) and (\ref{eq:3pt}). The reader should
  not confuse this notation with the notation for the Shapovalov form
  introduced earlier. Then, the OPEs take the structure
  \bea\label{eq:fope}
    & &\ww\!\!\!\!\Psi_{h_1;k_1}(z_1)\Psi_{h_2;k_2}(z_2)
       \ =\ \sum_h\sum_{k=0}^{r-1}
       \left(\prod_{i=0}^{r-1}\bra i,r-1-i\ket\right)^{-1} \\
    & &\ww\!\!\!\!\times\left|\begin{array}{ccccccc}
      \bra 0,0\ket  &\ldots&\bra   0,k-1\ket&\bra   0,k_1,k_2\ket&
          \bra   0,k+1\ket&\ldots&\bra   0,r-1\ket\\
      \vdots        &\ddots&\vdots          &\vdots              &
          \vdots          &\ddots&\vdots          \\
      \bra\ell,0\ket&\ldots&\bra\ell,k-1\ket&\bra\ell,k_1,k_2\ket&
          \bra\ell,k+1\ket&\ldots&\bra\ell,r-1\ket\\
      \vdots        &\ddots&\vdots          &\vdots              &
          \vdots          &\ddots&                \\
      \bra r-1,0\ket&\ldots&\bra r-1,k-1\ket&\bra r-s,k_1,k_2\ket&
          \bra r-1,k+1\ket&\ldots&\bra r-1,r-1\ket
    \end{array}\right|\Psi_{(h;k)}(z_2)\,,\nonumber
  \eea
  which in passing also proves that the matrix of two-point functions can be 
  inverted without problems. Of course, the denominator is written here in a 
  particularly symmetric
  way, it equals $\bra j,r-1-j\ket^r$ for any $0\leq j\leq r-1$. Note that the
  only non-zero entries above the anti-diagonal stem from the inserted column
  of three-point functions. The formula (\ref{eq:O=CD}) or (\ref{eq:fope}) 
  respectively are the sought after generalization of (\ref{eq:nope}) to the 
  case of logarithmic CFTs.

  With this result, we obtain in the simplest $r=2$ case the well known OPEs
  \begin{eqnarray}\label{eq:ope2a}
    \Psi_{(h_1;0)}(z)\Psi_{(h_2;0)}(0) &=& \sum_{h}
        \frac{C_{(h_1,h_2,h;1)}}{D_{(h,h;1)}^{}}\Psi_{(h;0)}(0)z^{h-h_1-h_2}
        \,,\\
    \label{eq:ope2b}\Psi_{(h_1;0)}(z)\Psi_{(h_2;1)}(0) &=& \sum_{h}
        \left[\frac{C_{(h_1,h_2,h;1)}}{D_{(h,h;1)}^{}}\Psi_{(h;1)}(0)\right.\\
    &+& \left.\frac{D_{(h,h;1)}C_{(h_1,h_2,h;2)}
           -D_{(h,h;2)}C_{(h_1,h_2,h;1)}}{D_{(h,h;1)}^2}\Psi_{(h;0)}(0)\right]
           z^{h-h_1-h_2}\,,\nonumber\\
    \label{eq:ope2c}\Psi_{(h_1;1)}(z)\Psi_{(h_2;1)}(0) &=& \sum_{h}
        \left[\left(\frac{C_{(h_1,h_2,h;2)}}{
           D_{(h,h;1)}^{}}
           - \frac{2C_{(h_1,h_2,h;1)}}{
           D_{(h,h;1)}^{}}\log(z)\right)\Psi_{(h;1)}(0)\right.\\
    &+& \left(\frac{D_{(h,h;1)}C_{(h_1,h_2,h;3)}
           -D_{(h,h;2)}C_{(h_1,h_2,h;2)}}{D_{(h,h;1)}^2}\right.
           \nonumber\\
    & & {}+\frac{2D_{(h,h;2)}C_{(h_1,h_2,h;1)}
           -D_{(h,h;1)}C_{(h_1,h_2,h;2)}}{D_{(h,h;1)}^2}\log(z)
           \nonumber\\
    & & \left.\left.{}-\frac{D_{(h,h;1)}C_{(h_1,h_2,h;1)}}{
           D_{(h,h;1)}^2}\log^2(z)\right)\Psi_{(h;0)}(0)
    \right]z^{h-h_1-h_2}\nonumber
  \end{eqnarray}
  Note that, for instance, the OPE of a proper primary with its
  logarithmic partner necessarily receives two contributions. One might
  naively have expected that proper primary fields do not change the
  J-level, although already the OPE of the stress-energy tensor with
  a logarithmic field will have an additional term involving the primary
  field. At the end of this section
  we will give a complete non-trivial example, namely the
  full set of generic OPE forms for a LCFT with rank four Jordan cells.

  But before doing so, we want to remark on the question of locality.
  The two- and three-point functions and the OPEs can easily be brought
  into a form for a local LCFT constructed out of left- and right-chiral
  half. The rule for this is simply to replace each $\log(z_{ij})$ by
  $\log|z_{ij}|^2$, and to replace each power $(z_{ij})^{\mu_{ij}}$ by
  $|z_{ij}|^{2\mu_{ij}}$. This yields a LCFT where all fields have the
  same holomorphic and anti-holomorphic scaling dimensions and the same
  J-level. Such an ansatz automatically satisfies both, the holomorphic
  as well as the anti-holomorphic Ward identities, if $z$ and $\bar z$
  are formally treated as independent variables. It is important to note,
  however, that the resulting full amplitudes do not factorize into
  holomorphic and anti-holomorphic parts. This is a well known feature of
  LCFTs. For example, the OPE equation (\ref{eq:ope2c}) would read
  in its full form
  \begin{eqnarray}\label{eq:ope3}
    \vv & & \Psi_{(h_1;1)}(z,\bar z)\Psi_{(h_2;1)}(0,0) = \sum_{h}
        |z|^{2(h-h_1-h_2)}
        \left[\frac{C_{(2)} - 2C_{(1)}\log|z|^2}{
           D_{(1)}^{}}\Psi_{(h;1)}(0,0)\right.\\
    \vv & & {}+\left(\left.\frac{D_{(1)}C_{(3)}
           -D_{(2)}C_{(2)}}{D_{(1)}^2}
        +\frac{2D_{(2)}C_{(1)}
           -D_{(1)}C_{(2)}}{D_{(1)}^2}\log|z|^2
        -\frac{D_{(1)}C_{(1)}}{
           D_{(1)}^2}\log^2|z|^2\right)\Psi_{(h;0)}(0,0)\right]\nonumber
  \end{eqnarray}
  with an obvious abbreviation for the structure constants. The reader
  is encouraged to convince herself of both, that on one hand this does indeed
  not factorize into holomorphic and anti-holomorphic parts, but that on the
  other hand this does satisfy the full set of conformal Ward identities.

  \subsection{{\sc A Non-trivial Example}}

   Now, we wish to present a fully worked out non-trivial example
   in order to demonstrate that even the generic structure of OPEs in
   arbitrary rank LCFTs is indeed more complicated than naively thought.
   Therefore, we present the OPEs for a rank four LCFT. Although all
   explicitly known LCFTs such as $c=-2$ and all the other $c_{p,1}$
   models \cite{Gurarie:1993,Flohr:1996a}, or certain non-trivial $c=0$ models
   \cite{Gurarie:1998,Gurarie:1999} are only rank two LCFTs, there are many
   indications that higher rank LCFTs exist. For instance, null-vectors
   for higher rank LCFTs have been noted in the latter two references in 
   \cite{Flohr:1996a}, and
   general considerations on higher rank LCFTs have been made in
   \cite{Ghezelbash:1997,Gravanis:2001}. Examples were found in the first
   work in \cite{Caux:1996} and in \cite{Maassarani:1997}.
   As a rule of thumb, one can reasonably conjecture
   that a CFT with a degenerate vacuum structure due to the existence
   of non-trivial zero-modes can be extended to a logarithmic CFT,
   whose maximal J-level (i.e. $r-1$) precisely equals the number of
   zero-modes (or of anti-commuting pairs of zero modes). 
   Again, $c=-2$ is here the prime example, since the
   well known $bc$ system of conformal spins 1 and 0 can indeed be
   expressed in terms of the $\theta_+\theta_-$ system briefly
   mentioned in section II.1. The $c=-2$ ghost system has one
   crucial zero-mode such that $\bra 0|0\ket = 0$, $\bra 0|c_0|0\ket \neq 0$.
   With the identification $c(z)=\theta_+(z)$, $b(z)=\partial\theta_-(z)$,
   the $\theta_+\theta_-$ system on one hand reproduces as a subset all the
   correlators of the $bc$ system when evaluated sandwiched between
   $\bra\xi_-|$ and $|0\ket$, and on the other hand constitutes an
   enlarged CFT which contains logarithmic fields. As discussed above,
   this CFT is logarithmic of rank two, as we would expect from the
   number of zero-modes. Work in this direction will appear
   elsewhere \cite{marco}.

   In order to
   keep the formulae readable, we will skip all the factors
   $(z_{12})^{h-h_1-h_2}$ as well as all arguments of the fields.
   Moreover, we only denote the J-levels in the structure constants
   omitting all references to the conformal weights. Hence, we put
   $C_k\equiv C_{(h_1,h_2,h;k)}$ and $D_k\equiv D_{(h_2,h;k)}$.
   Furthermore, $\ell^k$ is a shorthand for $\log^k(z_{12})$.
   \begin{eqnarray*}
     \Psi_{(h_1;0)}\Psi_{(h_2;0)} &=& \SU_h\frac{C_3}{D_3}\Psi_{(h;0)}\,,\\
     \Psi_{(h_1;0)}\Psi_{(h_2;1)} &=& \SU_h\frac{C_3}{D_3}\Psi_{(h;1)}
       + \frac{C_4D_3-C_3D_4}{D_3^2}\Psi_{(h;0)}\,,\\
     \Psi_{(h_1;0)}\Psi_{(h_2;2)} &=& \SU_h\frac{C_3}{D_3}\Psi_{(h;2)}
       + \frac{C_4D_3-C_3D_4}{D_3^2}\Psi_{(h;1)}
       + \frac{C_5D_3^2-C_4D_4D_3+C_3(D_4^2-D_5D_3)}{D_3^3}\Psi_{(h;0)}\,,\\
     \Psi_{(h_1;0)}\Psi_{(h_2;3)} &=& \SU_h\frac{C_3}{D_3}\Psi_{(h;3)}
       + \frac{C_4D_3-C_3D_4}{D_3^2}\Psi_{(h;2)}
       + \frac{C_5D_3^2-C_4D_4D_3+C_3(D_4^2-D_5D_3)}{D_3^3}\Psi_{(h;1)}\\
      &+&\frac{C_6D_3^3-C_5D_4D_3^2+C_4(D_4^2-D_5D_3)D_3-C_3(
         D_6D_3^2-2D_5D_4D_3+D_4^3)}{D_3^4}\Psi_{(h;0)}\,,\\
     \Psi_{(h_1;1)}\Psi_{(h_2;1)} &=& \SU_h\frac{C_3}{D_3}\Psi_{(h;2)}
       +\left(\frac{C_4D_3-C_3D_4}{D_3^2}-\frac{C_3}{D_3}\ell\right)
         \Psi_{(h;1)}\\
      &+&\left(\frac{C_5D_3^2-C_4D_4D_3+C_3(D_4^2-D_5D_3)}{D_3^3}
       - \frac{C_4D_3-C_3D_4}{D_3^2}\ell-\frac{C_3}{2D_3}\ell^2\right)
         \Psi_{(h;0)}\,,\\
     \Psi_{(h_1;1)}\Psi_{(h_2;2)} &=& \SU_h\frac{C_3}{D_3}\Psi_{(h;3)}
       + \left(\frac{C_4D_3-C_3D_4}{D_3^2}-\frac{C_3}{D_3}\ell\right)
         \Psi_{(h;2)}\\
      &+&\left(\frac{C_5D_3^2-C_4D_4D_3+C_3(D_4^2-D_5D_3)}{D_3^3}
       - \frac{C_4D_3-C_3D_4}{D_3^2}\ell\right)\Psi_{(h;1)}\\
      &+&\left(\frac{C_6D_3^3-C_5D_4D_3^2+C_4(D_4^2-D_5D_3)D_3
         -C_3(D_6D_3^2-2D_5D_4D_3+D_4^3)}{D_3^4}\right.\\
      & &\left.{\ }-\frac{C_5D_3^2-C_4D_4D_3+C_3(D_4^2-D_5D_3)}{D_3^3}\ell
       + \frac{C_3}{6D_3}\ell^3\right)\Psi_{(h;0)}\,,\\
     \Psi_{(h_1;1)}\Psi_{(h_2;3)} &=& \SU_h
         \left(\frac{C_4}{D_3}-\frac{2C_3}{D_3}\ell\right)\Psi_{(h;3)}
       + \left(\frac{C_5D_3-C_4D_4}{D_3^2}-\frac{C_4D_3-2C_3D_4}{D_3^2}\ell
         -\frac{C_3}{2D_3}\ell^2\right)\Psi_{(h;2)}\\
      &+&\left(\frac{C_6D_3^2-C_5D_4D_3+C_4(D_4^2-D_5D_3)}{D_3^3}\right.\\
      & &\left.{\ }-\frac{C_5D_3^2-C_4D_4D_3+2C_3(D_4^2-D_5D_3)}{D_3^3}\ell
       + \frac{C_3D_4}{2D_3^2}\ell^2 - \frac{C_3}{6D_3}\ell^3
         \right)\Psi_{(h;1)}\\
      &+&\left(\frac{C_7D_3^3-C_6D_4D_3^2+C_5(D_4^2-D_5D_3)D_3
         -C_4(D_6D_3^2-2D_5D_4D_3+D_4^3)}{D_3^4}\right.\\
      & &{\ }-\frac{C_6D_3^3-C_5D_4D_3^2+C_4(D_4^2-D_5D_3)
         -2C_3(D_6D_3^2-2D_5D_4D_3+D_4^3)}{D_3^4}\ell\\
      & &\left.{\ }-\frac{C_3(D_4^2-D_5D_3)}{2D_3^3}\ell^2
       + \frac{C_3D_4}{6D_3^2}\ell^3-\frac{C_3}{12D_3}\ell^4
         \right)\Psi_{(h;0)}\,,\\
     \Psi_{(h_1;2)}\Psi_{(h_2;2)} &=& \SU_h
         \left(\frac{C_4}{D_3} - \frac{2C_3}{D_3}\ell\right)\Psi_{(h;3)}
       + \left(\frac{C_5D_3-C_4D_4}{D_3^2}-\frac{C_4D_3-2C_3D_4}{D_3^2}\ell
         \right)\Psi_{(h;2)}\\
      &+&\left(\frac{C_6D_3^2-C_5D_4D_3+C_4(D_4^2-D_5D_3)}{D_3^3}\right.\\
      & &\left.{\ }-\frac{C_5D_3^2-C_4D_4D_3+2C_3(D_4^2-D_5D_3)}{D_3^3}\ell
       + \frac{C_4}{2D_3}\ell^2\right)\Psi_{(h;1)}\\
      &+&\left(\frac{C_7D_3^3-C_6D_4D_3^2+C_5(D_4^2-D_5D_3)D_3
         -C_4(D_6D_3^2-2D_5D_4D_3+D_4^3)}{D_3^4}\right.\\
      & &{\ }-\frac{C_6D_3^3-C_5D_4D_3^2+C_4(D_4^2-D_5D_3)
         -2C_3(D_6D_3^2-2D_5D_4D_3+D_4^3)}{D_3^4}\ell\\
      & &\left.{\ }+\frac{C_5D_3-C_4D_4}{2D_3^2}\ell^2
       + \frac{C_4}{6D_3}\ell^3-\frac{C_3}{12D_3}\ell^4
         \right)\Psi_{(h;0)}\,,\\
     \Psi_{(h_1;2)}\Psi_{(h_2;3)} &=& \SU_h
         \left(\frac{C_5}{D_3}-\frac{2C_4}{D_3}\ell+\frac{2C_3}{D_3}\ell^2
         \right)\Psi_{(h;3)}\\
      &+&\left(\frac{C_6D_3-C_5D_4}{D_3^2} - \frac{C_5D_3-2C_4D_4}{D_3^2}\ell
       - \frac{2C_3D_4}{D_3^2}\ell^2+\frac{5C_3}{6D_3}\ell^3\right)
         \Psi_{(h;2)}\\
      &+&\left(\frac{C_7D_3^2-C_6D_4D_3+D_5(D_4^2-D_5D_3)}{D_3^3}
       - \frac{C_6D_3^2-C_5D_4D_3+2C_4(D_4^2-D_5D_3)}{D_3^3}\ell\right.\\
      & &\left.{\ }+\frac{C_5D_3^2+4C_3(D_4^2-D_5D_3)}{2D_3^3}\ell^2
       - \frac{C_4D_3+5C_3D_4}{6D_3^2}\ell^3+\frac{C_3}{4D_3}\ell^4
         \right)\Psi_{(h;1)}\\
      &+&\left(\frac{C_8D_3^3-C_7D_4D_3^2+C_6(D_4^2-D_5D_3)D_3-C_5(
         D_6D_3^2-2D_5D_4D_3+D_4^3)}{D_3^4}\right.\\
      & &{\ }-\frac{C_7D_3^3-C_6D_4D_3^2+C_5(D_4^2-D_5D_3)D_3-2C_4(
         D_6D_3^2-2D_5D_4D_3+D_4^3)}{D_3^4}\ell\\
      & &{\ }+\frac{C_6D_3^3-C_5D_4D_3^2-4C_3(D_6D_3^2-2D_5D_4D_3+D_4^3)}{
         2D_3^4}\ell^2\\
      & &\left.{\ }+\frac{C_4D_4D_3+5C_3(D_4^2-D_5D_3)}{6D_3^3}\ell^3
       - \frac{C_4D_3+3C_3D_4}{12D_3^2}\ell^4+\frac{C_3}{12D_3}\ell^5
         \right)\Psi_{(h;0)}\,,\\
     \Psi_{(h_1;3)}\Psi_{(h_2;3)} &=& \SU_h
         \left(\frac{C_6}{D_3}-\frac{2C_5}{D_3}\ell+\frac{2C_4}{D_3}\ell^2
       - \frac{4C_3}{3D_3}\ell^3\right)\Psi_{(h;3)}
         +\left(\frac{C_7D_3-C_6D_4}{D_3^2}\right.\\
      & &\left.{\ }-\frac{C_6D_3-2C_5D_4}{D_3^2}\ell
       - \frac{2C_4D_4}{D_3^2}\ell^2+\frac{2C_4D_3+4C_3D_4}{3D_3^2}\ell^3
       - \frac{3C_3}{4D_3}\ell^4\right)\Psi_{(h;2)}\\
      &+&\left(\frac{C_8D_3^2-C_7D_4D_3+C_6(D_4^2-D_5D_3)}{D_3^3}
       - \frac{C_7D_3^2-C_6D_4D_3+2C_5(D_4^2-D_5D_3)}{D_3^3}\ell\right.\\
      & &{\ }+\frac{C_6D_3^2+4C_4(D_4^2-D_5D_3)}{2D_3^3}\ell^2
       - \frac{C_5D_3^2+2C_4D_4D_3+4C_3(D_4^2-D_3D_5)}{3D_3^3}\ell^3\\
      & &\left.{\ }+\frac{C_4D_3+3C_3D_4}{4D_3^2}\ell^4-\frac{C_3}{4D_3}\ell^5
         \right)\Psi_{(h;1)}\\
      &+&\left(\frac{C_9D_3^3-C_8D_3^2D_4+C_7(D_4^2-D_5D_3)D_3-C_6(
         D_6D_3^2-2D_5D_4D_3+D_4^2)}{D_3^4}\right.\\
      & &{\ }-\frac{C_8D_3^2-C_7D_4D_3^2+C_6(D_4^2-D_5D_3)D_3-2C_5(
         D_6D_3^2-2D_5D_4D_3+D_4^3)}{D_3^4}\ell\\
      & &{\ }+\frac{C_7D_3^3-C_6D_4D_3^2-4C_4(D_6D_3^2-2D_5D_4D_3+D_4^3)}{
         2D_3^4}\ell^2\\
      & &{\ }-\frac{C_6D_3^3-2C_5D_4D_3^2-4C_4(D_4^2-D_5D_3)-8C_3(
         D_6D_3^2-2D_5D_4D_3+D_4^3)}{6D_3^4}\ell^3\\
      & &\left.{\ }-\frac{C_5D_3^2+3C_4D_4D_3-9C_3(D_4^2-D_5D_3)}{12D_3^3}
         \ell^4
       + \frac{C_4D_3+3C_3D_4}{12D_3^2}\ell^5-\frac{5C_3}{72D_3}
         \ell^6\right)\Psi_{(h;0)}\,.
  \end{eqnarray*}
  Although this example seems tedious and lengthy, it is worth mentioning
  that it yields some surprises. For instance, the careful reader will note
  that the OPE $\Psi_{(h_1;1)}(z)\Psi_{(h_2;2)}(w)$ does not contain
  a term proportional to $\log(z_{12})^2$. Of course, it is clear from
  general arguments that this particular OPE may contain terms
  proportional to $\log(z_{12})^k$ for $k=0,1,2,3$, where 3 is the
  total J-level involved. The fact that the square term is missing
  is due to the general structure of the OPEs as required by global
  conformal covariance.

  It is illuminating to check the following: The three-point functions can
  be viewed as polynomials in the three variables $\ell=\log(z_{12})$,
  $\lambda=\log(z_{23})$ and $\Lambda=\log(z_{13})$. The two-point functions
  can then be seen as polynomials in the one variable $\lambda=\log(z_{23})$.
  Here, we again skip the trivial dependency on $\prod_{i<j}z_{ij}^{\mu_{ij}}$
  factors.
  The interesting fact, which also provides an excellent consistency check, is
  that the matrix product of $G^{(3)}_{k_1,k_2,k}(z_1,z_2,z_3)$ with
  the inverse of the matrix
  $G^{(2)}_{k,k_3}(z_2,z_3)$ yields sums of products of polynomials, namely
  $(G^{(3)}_{k_1}[\ell,\lambda,\Lambda])_{k_2,k}(G^{(2)}[\Lambda])^{k,k_3}
  =(C_{k_1+k_2}^{k_3}[\ell,\lambda,\Lambda])$,
  which always reduce for $\Lambda=\lambda$ to polynomials of the
  one variable $\ell$ only. To be more specific, the full set of
  two- and three-point functions as derived in this paper is indeed
  consistent with the above given OPE formula. In fact, since we have for the
  two-point functions
  \begin{equation}\label{eq:2pta}
    G^{(2)}_{k_2,k_3}[\lambda] =
    \delta_{h_2,h_3}\left(\sum_{m=0}^{k_2+k_3}D_{(h_2,h_3;k_2+k_3-m)}
    \frac{(-2)^{m}}{m!}\lambda^{m}\right){\rm e}^{-\lambda(h_2+h_3)}\,,
  \end{equation}
  and the three-point functions read
  \begin{eqnarray}\label{eq:3pta}
    & & G^{(3)}_{k_1,k_2,k_3}[\ell,\lambda,\Lambda]
    = \sum_{m=r-1}^{k_1+k_2+k_3}C_{(h_1,h_2,h_3;m)}
    \sum_{j_1=0}^{k_1}\sum_{j_2=0}^{k_2}\sum_{j_3=0}^{k_3}
    \delta_{j_1+j_2+j_3,k_1+k_2+k_3-m}\phantom{mmmn}\\
    & & \phantom{mmmm}\times\
    \frac{1}{j_1!j_2!j_3!}(\partial_{h_1})^{j_1}(\partial_{h_2})^{j_2}
    (\partial_{h_3})^{j_3}\left({\rm e}^{\ell(h_3-h_1-h_2)}
    {\rm e}^{\Lambda(h_2-h_1-h_3)}
    {\rm e}^{\lambda(h_1-h_2-h_3)}\right)\,,\nonumber
  \end{eqnarray}
  the matrix product for the computation of the OPE functions,
  \be
    \left.
    G^{(3)}_{k_1}[\ell,\lambda,\Lambda]\cdot\left(G^{(2)}[\lambda]\right)^{-1}
    \right|_{\Lambda=\lambda}
    = C_{k_1}[\ell]{\rm e}^{\ell(h_3-h_1-h_2)}\,,
  \ee
  yields a structure matrix with polynomial coefficients solely in the
  variable $\ell=\log(z_{12})$.
  
  In our example, we may for instance look at $\Psi_{(h_1;3)}\Psi_{(h_2;3)}$
  and there at the factor in front of the leading term $\Psi_{(h;3)}$ on the
  right hand side. This factor results from the sum of appropriate
  products of the following expressions (notation as above):
  \begin{eqnarray*}
    \bra\Psi_{(h_1;3)}\Psi_{(h_2;3)}\Psi_{(h;0)}\ket& =& 
      C_6 -2C_5\ell +2C_4\ell^2-\tfrac43C_3\ell^3\,,\\
    \bra\Psi_{(h_1;3)}\Psi_{(h_2;3)}\Psi_{(h;1)}\ket& =&
      C_7 -C_6(\lambda+\Lambda)-\tfrac{1}{12}C_3(\lambda-\Lambda)^4
      -(C_6-2C_5(\lambda+\Lambda))\ell\\
    &-& (\tfrac12C_3(\lambda-\Lambda)^2
      +2C_4(\lambda+\Lambda))\ell^2
      +\tfrac23(C_4+2C_3(\lambda+\Lambda))\ell^3-\tfrac34C_3\ell^4\,,\\
    \bra\Psi_{(h_1;3)}\Psi_{(h_2;3)}\Psi_{(h;2)}\ket& = &
      C_8-C_7(\lambda+\Lambda)+\tfrac12C_6(\lambda+\Lambda)^2
      -\tfrac{1}{12}C_4(\lambda-\Lambda)^4+\tfrac{1}{12}C_3(\lambda+\Lambda)
      (\lambda-\Lambda)^4\\
    &-& (C_7-C_6(\lambda+\Lambda)+C_5(\lambda+\Lambda)^2+\tfrac14
      C_3(\lambda-\Lambda)^4)\ell\\
    &+& \tfrac12(C_6+C_4(\lambda^2+\Lambda^2+6\lambda\Lambda)
      +C_3(\lambda+\Lambda)(\lambda-\Lambda)^2)\ell^2\\
    &-& \tfrac13(C_5+2C_4(\lambda+\Lambda)
      +\tfrac12C_3(5\lambda^2+5\Lambda^2+6\lambda\Lambda))\ell^3\\
    &+& \tfrac14(C_4+3C_3(\lambda+\Lambda))\ell^4
      -\tfrac14C_3\ell^5\,,\\
    \bra\Psi_{(h_1;3)}\Psi_{(h_2;3)}\Psi_{(h;3)}\ket& =&
      C_9-C_8(\lambda+\Lambda)+\tfrac12C_7(\lambda+\Lambda)^2-\tfrac16C_6
      (\lambda+\Lambda)^3-\tfrac{1}{12}C_5(\lambda-\Lambda)^4\\
    &+& \tfrac{1}{12}C_4(\lambda+\Lambda)(\lambda-\Lambda)^4
      -\tfrac{1}{72}(5\lambda^2+5\Lambda^2+2\lambda\Lambda)
      (\lambda+\Lambda)^4\\
    \lefteqn{\ww\ww\ww\ww\ww\ww-\ \ (C_8-C_7(\lambda+\Lambda)
      +\tfrac12C_6(\lambda+\Lambda)^2
      -\tfrac13C_5(\lambda+\Lambda)^3+\tfrac14C_4(\lambda-\Lambda)^4
      -\tfrac14C_3(\lambda+\Lambda)(\lambda-\Lambda)^4)\ell}\\
    \lefteqn{\ww\ww\ww\ww\ww\ww+\ \ \tfrac12(C_7-C_6(\lambda+\Lambda)
      -C_5(\lambda-\Lambda)^2
      +\tfrac13C_4(\lambda^2+\Lambda^2-10\lambda\Lambda)(\lambda+\Lambda)}\\
    &-& \tfrac14C_3(3\lambda^2+3\Lambda^2+2\lambda\Lambda)
      (\lambda-\Lambda)^2)\ell^2\\
    \lefteqn{\ww\ww\ww\ww\ww\ww-\ \ \tfrac{1}{18}(3C_6-6C_5(\lambda+\Lambda)
      -3C_4(\lambda^2+\Lambda^2+6\lambda\Lambda
      -C_3(7\lambda^2+7\Lambda^2+2\lambda\Lambda)
      (\lambda+\Lambda))\ell^3}\\
    \lefteqn{\ww\ww\ww\ww\ww\ww-\ \ \tfrac14(\tfrac13C_5
      +C_4(\lambda+\Lambda))\ell^4
      +\tfrac32C_3(\lambda+\Lambda)^2
      +\tfrac{1}{12}(C_4+3C_3(\lambda+\Lambda))\ell^5-\tfrac{5}{72}C_3\ell^6}
  \end{eqnarray*}
  for the three-point functions. Note that these expressions are all
  symmetric under the exchange $\lambda\leftrightarrow\Lambda$ as they
  should be. One needs also the two-point functions for which we have
  \begin{eqnarray*}
    \bra\Psi_{(h;0)}\Psi_{(h;3)}\ket &=& D_3\,,\\
    \bra\Psi_{(h;1)}\Psi_{(h;3)}\ket &=& D_4-2D_3\lambda\,,\\
    \bra\Psi_{(h;2)}\Psi_{(h;3)}\ket &=& D_5-2D_4\lambda+2D_3\lambda^2\,,\\
    \bra\Psi_{(h;3)}\Psi_{(h;3)}\ket &=& D_6-2D_5\lambda+2D_4\lambda^2
      -\tfrac43D_3\lambda^3\,.\\
  \end{eqnarray*}

  \section{{\sc Non-Proper Primary Fields}}

  {\sc We now turn} to the next complicated case, where we explicitly allow
  that the OPE of two primary fields might yield a logarithmic partner
  field on the right hand side. It is well known that such primary fields
  exist, in particular the so-called pre-logarithmic fields. In the
  $c=-2$ theory, the field $\mu$ with conformal weight $h=-1/8$ is such 
  a field, since it is a Virasoro primary field with OPE
  $\mu(z)\mu(w) = (z-w)^{1/4}(\tilde{\mathbb{I}}(w) - 2\log(z-w)\mathbb{I})$.
  However, the primary field $\mu$ is not itself part of a Jordan cell.
  The $c=-2$ theory provides another example of primary fields whose
  OPE also yields the field $\tilde{\mathbb{I}}$, namely the so-called
  fermionic fields $\theta_{\alpha}$. 
  With our notation introduced in section II.1, the
  former pre-logarithmic fields are twist fields, i.e.\ fields with
  non-trivial boundary conditions. Such fields do not have a
  zero mode content. The latter fields, however, have a zero mode
  content with the property that $Z_+(\theta_{\alpha})\neq Z_-(\theta_{
  \alpha})$, since $Z_{\beta}(\theta_{\alpha})=\delta_{\alpha\beta}$.
  These fields are again not members of Jordan cells.

  Of course, one might imagine a situation where non-proper primary
  fields do form part of a Jordan cell. The problem then is, that
  it is no longer possible to solve the conformal Ward identities in
  a hierarchical manner, as in section III, without further knowledge
  about the operator algebra. Basically, our approach in the third 
  section was to find the first non-vanishing two- and three-point
  functions, the ones with minimal zero mode content, and to derive
  correlators with higher zero-mode content by solving the inhomogeneous
  Ward identities step by step. If the zero mode content of all fields
  is known, we can estimate which two- and three-point functions might
  be non-zero, since the OPE must satisfy the bound
  \be
    Z_0(\Psi_3) \leq Z_0(\Psi_1) + Z_0(\Psi_2)
  \ee
  for all fields $\Psi_3(w)$ in the operator product of $\Psi_1(z)$ with
  $\Psi_2(w)$. This would provide us with a starting point for the
  hierarchical solution scheme.
  Unfortunately, this does not work for the twist fields,
  because no zero mode content can be defined for them. There is so far
  no example known where twist fields form part of a Jordan cell. We don't
  see an easy way to extend our description of Jordan cells in terms of
  zero mode content of fields such that it would encompass twist fields
  within Jordan cells. Thus
  we leave investigation of this case for future work. 

  \subsection{{\sc Fermionic Fields}}

  Let us now concentrate on the best known case of a rank $r=2$ LCFT, i.e.\
  where the maximal rank of Jordan cells is two, as in the prime example of
  the $c=-2$ theory. There, logarithmic
  operators, which together with their proper primary partners span the
  Jordan cells, are created by the operator $\tilde{\mathbb{I}}(z) =
  \Psi_{(0;1)}(z)$. As long as no twist fields are considered, we can construct
  all fields in terms of the pair of anti-commuting scalar $\theta$ fields
  defined in section II.1. Remember that the $\xi$ modes
  become the creation operators for logarithmic states. It is
  easy to see that proper primary fields do not possess any of the $\xi$ zero
  modes, while logarithmic fields possess precisely the zero mode contribution
  $\frac12\varepsilon^{\alpha\beta}\xi_{\alpha}\xi_{\beta}$. Since
  the $\xi$ modes do anti-commute, we call fields with just one $\xi$
  zero mode fermionic, and fields which are quadratic in $\xi$ bosonic.
  This coincides with the fact that for $c=-2$ all logarithmic fields and
  all proper primary fields have integer conformal weights. However,
  nothing prohibits us from considering the fields $\theta_{\alpha}(z)$
  themselves which also have zero conformal weight, but are fermionic.
  Many of the above arguments remain valid when we consider correlation
  functions involving $\theta$ fields. A further restriction is that
  the total number of $\theta$ fields must be even, since otherwise the
  correlation function vanishes identically. The reason is that consistency
  with the anti-commutation relations enforces to put $\bra\xi_+\ket =
  \bra\xi_-\ket = 0$. Only when the total number of $\theta$ fields is even,
  do we have a chance that a term $\xi_+\xi_-$ will survive after contraction.
  Moreover, the number of $\theta_+$ and $\theta_-$ fields must be equal, since
  otherwise $\theta_{\pm,0}$ zero modes will survive.

  More generally, a correlator of fields 
  $\bra\Psi_1(z_1)\ldots\Psi_n(z_n)\ket$ can only be non-zero, if it
  satisfies the conditions
  \be
  \sum_i Z_+(\Psi_i) = \sum_i Z_-(\Psi_i)\ \ \ \
  {\rm and}\ \ \ \ \sum_i Z_0(\Psi_i)\geq 2
  \ee
  in the rank two case. This statement is valid for fields $\Psi_i$ which
  are either proper primaries, logarithmic partners or fermionic fields (i.e.\
  fields with fermionic zero mode content).

  Correlation functions involving fermionic fields can be computed along
  the same lines as set out above. The only difference is that the
  action of the Virasoro algebra on fermionic fields does not have an
  off-diagonal part in the rank two case. More precisely, this is true for
  the $L_0$ mode of the Virasoro algebra, since this mode reduces always
  the total zero mode content symmetrically, i.e.\ it reduces $Z_-$ by the
  same amount as $Z_+$. Other modes, such as $L_1$, can reduce the zero
  mode content unevenly, as we will see in the next section. As long as
  we still assume that all fields in a Jordan cell are quasi-primary, we
  don't have to worry about this possibility. Of course, in LCFTs of
  higher rank, fermionic fields can easily admit an off-diagonal action
  of $L_0$, since $Z_{\pm}$ can both be larger than one. To keep the 
  following formul\ae\ reasonable simple, we won't consider this case here.

  The important fact is that the OPE of two fermionic fields produces
  a logarithm, i.e.
  \begin{equation}
    \theta_{\alpha}(z)\theta_{\beta}(0) = \varepsilon_{\alpha\beta}\left(
    \tilde{\mathbb{I}}(0) + (1+\log z)\mathbb{I}(0)\right)\,.
  \end{equation}
  This follows on general grounds, since $\bra\theta_{\alpha}(z)\theta_{\beta}
  (w)\ket=\varepsilon_{\alpha\beta}$ such that a three-point function of
  two fermionic and one logarithmic field necessarily involves a logarithm.
  The argument remains valid in the general rank two case and fields of
  arbitrary scaling dimension. Each Jordan cell is extended by two
  fermionic sectors such that we have the four fields
  $\Psi_{(h;0)}$, $\Psi_{(h;1)}$, and $\Psi_{(h;\pm)}$. It is then an easy
  task to compute all their OPEs from the two- and three-point functions
  \begin{eqnarray*}
    \bra\Psi_{(h;+)}(z_1)\Psi_{(h;-)}(z_2)\ket &=& \varepsilon_{+-}D_{(h,h;\pm)}
      (z_{12})^{-2h}\,,\\
    \bra\Psi_{(h;0)}(z_1)\Psi_{(h;1)}(z_2)\ket &=& D_{(h,h;1)}
      (z_{12})^{-2h}\,,\\
    \bra\Psi_{(h;1)}(z_1)\Psi_{(h;1)}(z_2)\ket &=& (D_{(h,h;2)}
      - 2D_{(h,h;1)}\log z_{12})(z_{12})^{-2h}\,,\\
    \bra\Psi_{(h_1;0)}(z_1)\Psi_{(h_2;0)}(z_2)\Psi_{(h_3;1)}(z_3)\ket &=&
      C_{(h_1,h_2,h_3;1)}
      (z_{12})^{h_3-h_1-h_2}(z_{13})^{h_2-h_1-h_3}(z_{23})^{h_1-h_2-h_3}\,,\\
    \bra\Psi_{(h_1;0)}(z_1)\Psi_{(h_2;+)}(z_2)\Psi_{(h_3;-)}(z_3)\ket &=&
      \varepsilon_{+-}C_{(h_1;0)(h_2,h_3;\pm)}
      \PI(z_{ij})^{h_k-h_i-h_j}\,,\\
    \bra\Psi_{(h_1;0)}(z_1)\Psi_{(h_2;1)}(z_2)\Psi_{(h_3;1)}(z_3)\ket &=&
      (C_{(h_1,h_2,h_3;2)} - 2C_{(h_1,h_2,h_3;1)}\log z_{23})
      \PI(z_{ij})^{h_k-h_i-h_j}\,,\\
    \bra\Psi_{(h_1;1)}(z_1)\Psi_{(h_2;+)}(z_2)\Psi_{(h_3;-)}(z_3)\ket &=&
      \varepsilon_{+-}(
      C_{(h_1;0)(h_2,h_3;\pm)}(\log z_{23} - \log z_{12} - \log z_{13}) \\
      &+&C_{(h_1;1)(h_2,h_3;\pm)})
      \PI(z_{ij})^{h_k-h_i-h_j}\,,\\
    \bra\Psi_{(h_1;1)}(z_1)\Psi_{(h_2;1)}(z_2)\Psi_{(h_3;1)}(z_3)\ket &=&
      (C_{(h_1,h_2,h_3;3)} - C_{(h_1,h_2,h_3;2)}(\log z_{12} + \log z_{13}
      + \log z_{23})\\
      &+&2C_{(h_1,h_2,h_3;1)}(\log z_{12}\log z_{13}
      + \log z_{12}\log z_{23} + \log z_{13}\log z_{23}\\
      & &{} - \tfrac12\log^2 z_{12}
      - \tfrac12\log^2 z_{13} - \tfrac12\log^2 z_{23}))
      \PI(z_{ij})^{h_k-h_i-h_j}\,,
  \end{eqnarray*}
  and permutations. Note that we have explicitly indicated the antisymmetry
  under exchanging the order of the fermionic fields. These results
  agree in the special case where all $h_i=0$
  with the explicit calculations for the $c=-2$ LCFT by Kausch
  \cite{Kausch:1995}. The singular terms of the corresponding additional 
  OPEs read
  \bea
     \Psi_{(h_1;0)}(z_1)\Psi_{(h_2;\pm)}(z_2) &=& \sum_h
     \frac{C_{(h_1;0)(h_2,h;\pm)}}{D_{(h_2,h;\pm)}}(z_{12})^{h-h_1-h_2}
       \Psi_{(h;\pm)}(z_2)\,,\\
     \Psi_{(h_1;\pm)}(z_1)\Psi_{(h_2;\mp)}(z_2) &=& \sum_h
       \varepsilon_{\mp\pm}\left[\left(
         \frac{C_{(h_1,h_2;\pm)(h;0)}D_{(h_2,h;2)} -
         C_{(h_1,h_2;\pm)(h;1)}D_{(h_2,h;1)}}{D_{(h_2,h;1)}^2}\right.\right.
         \nonumber\\
     \lefteqn{\vv\vv\!\!\!\left.\left.-
         \frac{C_{(h_1,h_2;\pm)(h;0)}}{D^{}_{(h_2,h;1)}}\log(z_{12})\right)
         \Psi_{(h;0)}(z_2) +
         \frac{C_{(h_1,h_2;\pm)(h;0)}}{D^{}_{(h_2,h;1)}}\Psi_{(h;1)}(z_2)
         \right](z_{12})^{h-h_1-h_2}\,,\nonumber}\\
     \Psi_{(h_1;1)}(z_1)\Psi_{(h_2;\pm)}(z_2) &=& \sum_h
       \frac{C_{(h_1;1)(h_2,h;\pm)}
         -C_{(h_1;0)(h_2,h,\pm)}\log(z_{12})}{D_{(h_2,h;\pm)}}
       (z_{12})^{h-h_1-h_2}\Psi_{(h;\pm)}(z_2)\,.\nonumber
  \eea
  The above statement shows that rank two LCFTs naturally allow for
  fermionic fields. It has been suggested in \cite{MoghimiAraghi:2001a} 
  to formally collect
  these quadruples of fields in ``superfields'' $\Psi_h(z,\eta^+,\eta^-)$ of
  $N\!=\!2$ Grassmann variables such that
  \begin{equation}
    \Psi_h(z,\eta^+,\eta^-) = \Psi_{(h;0)}(z) + \eta^+\Psi_{(h;-)}(z)
    + \eta^-\Psi_{(h;+)}(z) + \eta^+\eta^-\Psi_{(h;1)}(z)\,,
  \end{equation}
  which in the $c=-2$ case resembles the $\xi_{\pm}$ zero mode contributions.
  It is tempting to conjecture that a rank $k$ LCFT
  will naturally incorporate the analog of anti-commuting scalars
  for $\mathbb{Z}_k$ para-fermions, whose OPEs among them create logarithmic
  fields of according J-levels. However, an investigation of this is
  beyond the scope of the present paper.

  \subsection{{\sc Twist Fields}}

  As already explained, there is (at least) one more sort of fields which 
  may occur in LCFTs.
  In the standard $c=-2$ example, the two fields $\mu(z)$ and $\sigma(z)$
  with conformal weights $h_{\mu}=-1/8$ and $h_{\sigma}=3/8$ respectively,
  are not yet accounted for. These fields are twist fields. They can be
  treated much along the same lines as fermionic fields. The difference
  is that their mode expansion is in $\mathbb{Z}+\iota$ with a certain
  rational $\iota$ depending on the boundary conditions and the
  ramification number of the twists.
  The fields $\mu$ and $\sigma$ are $\mathbb{Z}_2$ twists.
  Despite the difference in the mode expansion, twist fields behave
  quite similar to the (para-)fermionic fields mentioned above.
  In particular, their two-point functions are non-zero if and only if
  they involve a twist $\chi_{\iota}$ and its anti-twist $\chi_{\iota^*}$,
  which resembles the fact that for fermionic fields only the
  two-point function of two different fermions is non-zero.
  Higher twist fields are then analogous to para-fermions. 

  One might attempt to extend the definition of zero mode content
  to rational numbers such that a twist field with mode expansion
  in $\mathbb{Z}+\iota$ would get assigned $Z_0(\chi_{\iota})=
  Z_+(\chi_{\iota})+Z_-(\chi_{\iota}) = \iota+\iota^*$.
  A necessary condition for a correlator $\bra\chi_{\iota_1}\ldots
  \chi_{\iota_n}\ket$ to be non-zero would then read
  \be\label{eq:zmc}
    \sum_iZ_+(\chi_{\iota_i})\in \mathbb{Z}\ \ \ \ {\rm and}\ \ \ \ 
    \sum_iZ_-(\chi_{\iota_i})\in \mathbb{Z}\,.
  \ee
  However, such an assignment is obviously only determined modulo integers,
  and it is not a priory clear how to implement the condition for a
  minimal zero mode content in general. At least, 
  one should now always consider separately the zero mode content
  $Z_+$ and $Z_-$, as indicated above. In the rank two case, however, we can
  incorporate the condition for a minimal zero mode content into
  (\ref{eq:zmc}) by simply replacing $\mathbb{Z}$ by $\mathbb{N}$.
  Indeed, the $\mathbb{Z}_2$ twist fields
  $\mu$ and $\sigma$ have twist numbers $(\iota,\iota^*) = (\frac12,\frac12)$
  and $(-\frac12,\frac32)$ respectively, from which we immediately can
  read off, which two- and three-point functions of $\mu$ and $\sigma$ fields
  can be non-zero. 

  To emphasize the common features of fermionic and twist fields,
  we contrast their possible two- and three-point functions with the
  ones for fermionic fields (there are no non-vanishing two- or
  three-point functions involving both, fermionic and twist fields,
  simultaneously). The
  notation $\iota^*$ means the anti-twist $1-\iota$ with respect to $\iota$,
  and one always has $h_{\iota} = h_{\iota^*}$.
  The only nontrivial two-point function then reads
  \begin{equation}
    \bra\chi_{\iota}(z_1)\chi_{\iota^*}(z_2)\ket =
    D_{\iota\iota^*}(z_{12})^{-2h_{\iota}}
  \end{equation}
  with $D_{\iota\iota^*}=D_{\iota^*\iota}$. Note that in contrast to
  the fermionic fields, twist fields are symmetric.
  The three-point functions are easily computed and the results are
  \begin{eqnarray*}
    \bra\Psi_{(h_1;0)}\Psi_{(h_2;k')}\chi_{\iota_3}\ket &=& 0 \,, \\
    \bra\Psi_{(h_1;0)}\chi_{\iota_2}\chi_{\iota_3}\ket &=&
      \delta_{\iota^{}_3,\iota_2^*}
      C_{(h_1,0)\iota^{}_2\iota_3^*}\PI(z_{ij})^{h_k-h_i-h_j}\,, \\
    \bra\Psi_{(h_1;1)}\chi_{\iota_2}\chi_{\iota_3}\ket &=&
      \delta_{\iota^{}_3,\iota_2^*}
      (C_{(h_1,1)\iota^{}_2\iota_3^*} + C_{(h_1,0)\iota^{}_2\iota_3^*}
      (\log z_{23} - \log z_{12} - \log z_{13}))\PI(z_{ij})^{h_k-h_i-h_j}\,,\\
    \bra\chi_{\iota_1}\chi_{\iota_2}\chi_{\iota_3}\ket &=&
      (\delta_{\iota^{}_3,\iota_1^*+\iota_2^*}
      + \delta_{\iota^{}_3,1-\iota^{}_1-\iota^{}_2})
      C_{\iota_1\iota_2\iota_3}\PI(z_{ij})^{h_k-h_i-h_j} \,, \\
    \bra\Psi_{(h_1;1)}\Psi_{(h_2;1)}\chi_{\iota_3}\ket &=&
      C_{(h_1,h_2;2)\iota^{}_3}
      \PI(z_{ij})^{h_k-h_i-h_j}\,.
  \end{eqnarray*}
  Note that some of the introduced constants may be zero, e.g.\
  $C_{\iota_1\iota_2\iota_3} = 0$ whenever the three twists do not
  add up to an integer. Most remarkably is
  perhaps the fact that $\bra\Psi_{(h_1;1)}\Psi_{(h_2;1)}\chi_{\iota_3}\ket$
  might be non-zero. This does not happen in the $c=-2$ theory, since it
  implies that the OPE of two logarithmic fields has a contribution
  \begin{equation}
    \Psi_{(h;1)}(z)\Psi_{(h';1)}(0) = \ldots + \frac{C_{(h,h';2)\iota}}{
      D_{\iota\iota^*}}z^{h_{\iota}-h-h'}\chi_{\iota}(0) + \ldots\,,
  \end{equation}
  which is not the case in the $c=-2$ theory. However, already the next
  theory in the $c_{p,1}$ series of LCFTs, namely the $c_{3,1}=-7$ model,
  shows precisely this feature, where the fusion rule of the $h=0$
  logarithmic field with itself involves the twist field with $h=-\frac13$
  on the right hand side. Since the main focus of this paper lies on
  logarithmic fields, we will not go into further detail here. The
  OPEs involving $\chi_{\iota}$ fields read correspondingly
  \begin{eqnarray*}
    \Psi_{(h;0)}(z)\chi_{\iota}(0) &=& \frac{C_{(h;0)\iota\iota'}}{
      D_{\iota\iota^*}}z^{h_{\iota'}-h_{\iota}-h}\chi_{\iota'}(0)
      \,,\\
    \Psi_{(h;1)}(z)\chi_{\iota}(0) &=& \frac{C_{(h,h';2)\iota}}{
      D_{(h;1)}}z^{h'-h-h_{\iota}}\Psi_{(h';0)}\\
      &+&
      \frac{C_{(h;1)\iota\iota'}-C_{(h;0)\iota\iota'}\log z}{
      D_{\iota\iota^*}}z^{h_{\iota'}-h_{\iota}-h}\chi_{\iota'}(0) +
      \frac{C_{(h;1)\iota{\iota'}^*}}{D_{\iota\iota^*}}z^{h_{{\iota'}^*}-
      h_{\iota}-h}\chi_{{\iota'}^*}(0)\,,\\
    \chi_{\iota}(z)\chi_{\iota'}(0) &=& \frac{C_{(h;1)\iota\iota'}}{
      D_{(h;1)}}z^{h'-h_{\iota}-h_{\iota'}}\Psi_{(h;0)}(0) +
      \frac{C_{\iota\iota'\iota''}}{D_{\iota\iota^*}}z^{h_{\iota''}-h_{\iota}
      -h_{\iota'}}\chi_{\iota''}(0)\,,\\
    \chi_{\iota}(z)\chi_{\iota^*}(0) &=& \frac{C_{(h;1)\iota\iota^*}D_{(h;1)}
      - C_{(h;0)\iota\iota^*}D_{(h;2)} + C_{(h;0)\iota\iota^*}D_{(h;1)}\log z}{
      D_{(h;1)}^2}z^{h-h_{\iota}-h_{\iota^*}}\Psi_{(h;0)}(0)\\
      &+&
      \frac{C_{(h;0)\iota\iota^*}}{D_{(h;1)}}z^{h-h_{\iota}-h_{\iota^*}}
      \Psi_{(h;1)}(0) \,,
  \end{eqnarray*}
  where in the last two equations $\iota'\neq\iota^*$.
  As remarked above, some of the structure constants may vanish, as they
  do in the $c=-2$ LCFT. One sees that even the simple rank two case
  gets quite complicated and needs a cumbersome notation. The situation is
  slightly better in the particular case for the $c=-2$ theory where all
  amplitudes involving up to four twist fields as well as amplitudes with an
  arbitrary number of fermionic fields were computed in \cite{Kausch:1995}.

\section{\reseteqn{\sc Non-quasi-primary fields}}

  {\sc Our discussion of} correlation functions and operator product 
  expansions in logarithmic CFTs heavily relies on the following
  assumption which we so far have made: that all logarithmic
  partner fields within a Jordan cell be quasi-primary. This means in
  particular that $L_1|h;k\ket = L_1\Phi_{(h;k)}(0)|0\ket = 0$. As a 
  consequence, we could make elaborate use of the Ward identities of
  global conformal transformations in the form (\ref{eq:ward}).
  This section is devoted to the question under which more relaxed
  circumstances our results still hold.

  It is by no means clear that logarithmic partner fields are indeed all
  quasi-primary. On the contrary,
  even the simplest known LCFT, the $c=-2$ model, features a Jordan cell
  where the logarithmic partner is not quasi-primary \cite{Gaberdiel:1996a}. 
  Actually, as already outlined in section II, in this
  model exists a Jordan cell for conformal weight $h=1$, built from
  a primary field\footnote{More precisely, we are dealing with a doublet of
  two such fields, distinguished by the $\alpha$-label.}
  $\partial\theta_{\alpha}(z)$ and its logarithmic
  partner field $\mbox{:$\theta_+\theta_-\partial\theta_{\alpha}$:}(z)$.
  Here, we again used the realization of the $c=-2$ theory in terms of
  two anti-commuting scalar fields $\theta_{\pm}(z)$ along the lines
  (\ref{eq:theta}) and (\ref{eq:thth}). It is easy to see that 
  $|1;0\ket=\theta_{\alpha,-1}|0\ket$
  and that $|1;1\ket=\theta_{\alpha,-1}(\xi_+\xi_- + 1)|0\ket$. It follows
  that $L_1|1;0\ket = 0$, while $L_1|1;1\ket = -\xi_{\alpha}|0\ket\neq 0$,
  where one uses that the stress-energy tensor is given by (\ref{eq:T}).
  We remind the reader that $\xi_{\alpha}$ is one of the two zero-modes
  of the field $\theta_{\alpha}(z)$. Hence,
  $|1;1\ket$ is not a state corresponding to a quasi-primary field.

  The point is that it does not matter. The global Ward identities are
  not affected by this non-zero term. More generally, all correlation functions
  involving the field $\Psi_{(h=1;1)}(z)$, the logarithmic partner of the
  primary $h=1$ field $\Psi_{(h=1;0)}(z)=\partial\theta(z)$, 
  behave exactly as if the
  field were quasi-primary. The reason for this is simply that the
  state $L_1|1;1\ket$ is fermionic with respect to the number of
  $\xi_{\alpha}$ zero modes. More precisely, it has $Z_+=1$, $Z_-=0$ (or
  vice versa).
  Any correlation function can only be
  non-zero if the numbers $(Z_+,Z_-)$ of $\xi_{\pm}$ zero modes are,
  after all contractions are done, exactly $(1,1)$. Hence, any 
  correlation function involving $\Psi_{(h=1;1)}(z)$, which has a
  chance to be non-zero, must initially have fermion numbers $(Z_+,Z_-)$ 
  with $Z_++Z_-$ even and $Z_+\geq 1$, $Z_-\geq 1$. Applying $L_1$ to it 
  results in the global conformal Ward identity up to additional terms with
  fermion numbers $(Z_+-1,Z_-)$ or $(Z_+,Z_--1)$. Since $Z_++Z_--1$ is then 
  necessarily an odd number, the additional terms must vanish. It follows that
  the fact that $\Psi_{(h=1;1)}(z)$ is not quasi-primary does not
  influence the correlation functions, because the spoiling term does
  not lead to any non-vanishing contributions. Note that this 
  statement is only true as long as we consider the effect within
  correlation functions. The deeper reason is that the action of the
  Virasoro algebra changes the fermion numbers unevenly in this case.

  Of course, not all correlation functions involving a field
  $\theta_\alpha(z)$ corresponding to the state $\xi_\alpha|0\ket$ automatically
  vanish. For example, $\bra 0|\theta_+(z)\theta_-(w)|0\ket\neq 0$. 
  What is meant in the above discussion is that all correlation functions
  vanish, which result from applying $L_1$ to a correlator involving
  $\Psi_{(h=1;1)}(z)$ and other fields such that the initial correlator
  might be non-zero. Since $L_1$ acts as a derivation, it only changes
  one of the inserted fields at a time, so that starting with an admissible
  number of $\xi_\alpha$ zero modes leads to terms with non-admissible
  numbers of $\xi_\alpha$ zero modes.

  There are strong indications that this structure of Jordan cells with
  non-quasi-primary fields is more generally true.
  At least, there is so far no LCFT explicitly known where logarithmic partner
  fields are not quasi-primary in a way which would affect correlation
  functions and therefore our general conclusions on their general
  structure. All LCFTs which can be constructed or realized explicitly
  in terms of fundamental free fields, such as the $c=-2$ model, 
  receive their peculiar logarithmic fields ultimately due to the
  existence of conjugate pairs of zero-modes. In the $c=-2$ model,
  these are the two pairs $\xi_{\pm},\theta_{\mp,0}$ respectively (see
  sect.\ II.1).
  In this situation, we only need that the logarithmic partner fields
  are quasi-primary up to terms with the ``wrong'' number of such
  zero modes. Here, wrong means in the above explained sense that the
  number of zero modes gets changed unevenly.

  Following the lines of \cite{MoghimiAraghi:2001a},
  the effective zero mode content is equivalently described in
  terms of nilpotent variables with which the fields spanning a Jordan
  cell are grouped together in a superfield like fashion. In the rank two
  case this is easily accomplished by introducing for each Jordan cell
  two Grassmann variables
  $\eta^{\pm}$ and a bosonic nilpotent variable $\Theta=\eta^+\eta^-$ with
  the property $\Theta^2=0$. Then, we may group together all fields,
  the primary $\Psi_{(h;0)}(z)$, its logarithmic partner $\Psi_{(h;1)}(z)$
  and the two fermionic fields $\Psi_{(h;\pm)}(z)$ as
  {\boldmath $\Psi$}$_h(z)=\Psi_{(h;0)}(z) + \eta^+\Psi_{(h;+)}(z)
  +\eta^-\Psi_{(h;-)}(z) + \eta^+\eta^-\Psi_{(h;1)}(z)$.
  For higher rank Jordan cells, an analogous procedure applies. 
  For a Jordan cell of {\em even\/} rank $2r$
  one needs $r-1$ pairs of Grassmann variables. The member of
  J-level $k$ in the Jordan cell itself is given by the elementary
  completely symmetric 
  polynomial $\sigma_k(\Theta_1,\ldots,\Theta_{r-1})=
  \sigma_k(\eta_1^+\eta_1^-,\ldots,\eta_{r-1}^+\eta_{r-1}^-)$ of 
  total degree $2k$ in the $\eta$'s. So the primary field 
  $\Phi_{(h;0)}(z)$ belongs to 
  $\sigma_0\equiv 1$, while the top J-level field $\Phi_{(h;r-1)}(z)$ 
  belongs to $\sigma_{r-1}\propto
  \eta_1^+\eta_1^+\ldots\eta_{r-1}^+\eta_{r-1}^-$.
  The elementary completely symmetric polynomial $\sigma_k(x_1,\ldots,x_n)$
  is defined as $\sigma_k=\sum_{i_1<i_2<\ldots i_k}
  x_{i_1}x_{i_2}\ldots x_{i_k}$ up to normalization. 
  Other polynomials $p(\{\eta_i^{\pm}\})$,
  whose monomials $m_{\{i\}}=\prod_k\eta_{i_k}^{\alpha_k}$
  have differing partial degrees ${\rm deg}_{\eta_j^+}(m_{\{i\}})\neq
  {\rm deg}_{\eta_j^-}(m_{\{i\}})$ for at least one $j$, belong to fields
  which are the higher rank analogons of the fermionic fields described above.
  Overall symmetry in the Grassmann variables demands that all polynomials
  must be symmetric polynomials in the $\eta$'s. However, what may be
  varied is how the total degree is split into $\eta^+$ variables and
  $\eta^-$ variables. Hence, we use the symmetric polynomials in two
  colors, $\sigma_k^{l,k-l}$, instead. These can be written as
  $\sigma_k^{l,k-l}(\{\eta^+,\eta^-\}) = \sigma_{{\rm max}(l,k-l)}(\{\Theta\})
  \sigma_{{\rm min}(l,k-l)}(\{\eta_{\alpha_{{\rm min}(l,k-l)}}\})$. Note
  that $\sigma_k^{l,k-l}=0$ for $l$ or $k-l$ larger $r-1$.
  The Jordan cell is then spanned by
  the fields corresponding to the symmetric polynomials 
  $\sigma_{2k}^{k,k}(\{\eta^+,\eta^+\})=\sigma_k(\{\Theta\})$.

  For example, in the rank four case we have the following possibilities:
  The Jordan cell is spanned by
  \begin{eqnarray*}
    \mbox{\boldmath $\Phi$}_h(z)&=&
    \Phi_{(h;0)}(z) + 
    \sum_{1\leq i\leq 3}\eta_i^+\eta_i^-\,\Phi_{(h;1)}(z) +
    \sum_{1\leq i<j\leq 3}\eta_i^+\eta_i^-\eta_j^+\eta_j^-\,\Phi_{(h;2)}(z)\\
    & & \mbox{}+\
    \eta_1^+\eta_1^-\eta_2^+\eta_2^-\eta_3^+\eta_3^-\,\Phi_{(h;3)}(z)\\
    &=&
    \sigma_0\Phi_{(h;0)}(z)+\sigma_2^{1,1}\Phi_{(h;1)}(z)
    +\sigma_4^{2,2}\Phi_{(h;2)}(z)+\sigma_6^{3,3}\Phi_{(h;3)}(z)\,.
  \end{eqnarray*}
  Furthermore, we have symmetrized non-bosonic fields according to the
  following ``diamond'' of two-color symmetric polynomials:
  $$
    \begin{array}{ccccccc}
           &           &           &\si{0}{0}  &           &           &      \\
           &           &\si{1}{1,0}&           &\si{1}{0,1}&           &      \\
           &\si{2}{2,0}&           &\si{2}{1,1}&           &\si{2}{0,2}&      \\
\si{3}{3,0}&           &\si{3}{2,1}&           &\si{3}{1,2}&           &
\si{3}{0,3}\,.\\
           &\si{4}{3,1}&           &\si{4}{2,2}&           &\si{4}{1,3}&      \\
           &           &\si{5}{3,2}&           &\si{5}{2,3}&           &      \\
           &           &           &\si{6}{3,3}&           &           &
    \end{array}
  $$
  Of course, it is not clear whether all higher rank LCFTs fall into this
  pattern, but the crucial role of conjugate zero mode pairs suggests so.

  It is now easy to see, which correlation functions may be non-zero.
  Clearly, each inserted field $\Phi$ comes with an associated polynomial
  $\sigma$. Consequently, an arbitrary $n$-point function can only be 
  non-vanishing,
  if the product $\prod_{i=1}^n\sigma_{k_i}^{l_i,k_i-l_i}$ meets the
  conditions 
  \be\label{eq:si}
    \sum_{i=1}^n l_i = \sum_{i=1}^n (k_i-l_i)\geq r-1\,.
  \ee
  Of course, this is nothing else than our initial conditions on
  the zero mode content, since any field with associated polynomial
  $\sigma_{k}^{l,k-l}$  has zero mode content $Z_+=l$, $Z_-=k-l$.

  The action of symmetry generators such as the modes of the
  Virasoro algebra may have off-diagonal contributions meaning that they
  change a field into another and thus change a given associated polynomial
  into another. However, this off-diagonal term contributes to the
  correlator only if the resulting product of associated polynomials again
  satisfies the above condition (\ref{eq:si}). Now, symmetry generators
  always act as derivations on correlators such that an immediate necessary
  condition on an off-diagonal term is that the off-diagonal action of
  the generator moves vertically in the $\sigma$-diamond. Otherwise, the
  off-diagonal term has no effect in the correlator. Coming back to our initial
  example in the $c=-2$ theory, we have that $|1;1\ket$ has associated
  polynomial $\sigma_{2}^{1,1}$, while $L_1|1;1\ket$ has associated
  polynomial $\sigma_{1}^{1,0}$. Thus, $L_1$ does not move vertically in
  the $\sigma$-diamond, and the non-quasi-primary term hence does 
  not contribute.

  We therefore arrive at the following generalized picture:
  The action of symmetry generators such as the Virasoro algebra in a 
  logarithmic CFT possesses off-diagonal additional terms with accompanying
  moves in the associated $\sigma$-diamond. However, all such terms with
  a non-vertical move are irrelevant when considered within correlation
  functions. All moves in the $\sigma$-diamond are obviously generated
  by the two basic moves 
  \be
    Q^+:\sigma_k^{l,k-l}\mapsto\sigma_{k-1}^{l,k-1-l}
    \ \ \ \ {\rm and}\ \ \ \ 
    Q^-:\sigma_k^{l,k-l}\mapsto\sigma_{k-1}^{l-1,k-l}\,.
  \ee 
  The careful
  reader might notice that these moves always go upwards within the
  diamond. However, since symmetry generators are presumably free of
  zero modes which do not annihilate the vacuum (otherwise, the vacuum is
  no longer invariant under the considered symmetry), we do not expect
  that a symmetry generator will move downwards within the diamond. 
  These two basic moves may be viewed as generating a BRST like symmetry,
  since with $\bra\prod_i\sigma_{k_i}^{l_i,k_i-l_i}\ket\neq 0$, we
  certainly have 
  \be
    Q^{+}\bra\prod_i\sigma_{k_i}^{l_i,k_i-l_i}\ket=
    \sum_j\bra\sigma_{k_1}^{l_1,k_1-l_1}\ldots\sigma_{k_j-1}^{l_j,k_j-1-l_j}
    \ldots\sigma_{k_n}^{l_n,k_n-l_n}\ket = 0\,,
  \ee
  and analogously for $Q^-$.
  Actually, one may introduce operators $Q_{\ell}^{\pm}$ with $1\leq\ell\leq
  r-1$ which act by setting $\eta^{\mp}_{\ell}$ formally to one by contracting
  with a conjugate Grassmann variable, i.e.\ by acting with $\{
  \bar\eta_{\ell}^{\mp},\cdot\}$. These latter operators have the nice
  property to automatically satisfy $(Q^{\pm}_{\ell})^2=0$, while
  $Q^+_{\ell}Q^-_{\ell}\neq 0$. These latter operators may then indeed be
  considered as BRST symmetries on correlation functions. We
  introduce $Q_{\pm}$ as symbols for
  classes of such symmetries, denoting arbitrary moves upwards in the diamond,
  which are not vertical. Thus, these classes 
  contain also products such as $Q^+_{\ell_1}Q^+_{\ell_2}$ etc.

  We may thus finally generalize our assumptions on the fields in LCFTs
  such that the action of the Virasoro algebra only needs to satisfy our
  basic conditions (\ref{eq:vir}) and (\ref{eq:ward}) up to terms, which
  can be written as $Q_{\pm}$ of an admissible configuration, i.e.\
  can be written as $Q_{\pm}\bra\prod_i\sigma_{k_i}^{l_i,k_i-l_i}\ket$
  with $\bra\prod_i\sigma_{k_i}^{l_i,k_i-l_i}\ket$ satisfying (\ref{eq:si}).
  If this is the case, all our results on two- and three-point functions
  and on the OPE structure remain valid in precisely the form as we have
  given them. All explicitly known LCFTs seem to possess only Jordan cells
  (and associated non-bosonic fields) which fall into this pattern.
  However, we do not have a proof that this has always to be the case.
  The strongest indications in favor of such a conjecture stem from the
  classification of possible Jordan structure like modules in the
  representation theory of the Virasoro algebra as given in 
  \cite{Rohsiepe:1996}. A deeper study, which also should consider Jordan cell
  structures with respect to other chiral symmetries has to be left for
  future work.

  Last but not least, we remark that the zero mode content further implies
  (\ref{zmc}) and hence (\ref{jlevel}). Let us momentarily assume that
  these bounds would not be strict. Then, for example, a two-point function
  $\bra\Psi_{(h;0)}(z)\Psi_{(h;0)}(w)\ket$ could be non-zero, if the OPE
  of the primary field $\Psi_{(h;0)}$ with itself contained
  $\Psi_{(0;r-1)}$ in the rank $r$ case. 
  It is an easy task to check that this is, however,
  inconsistent, since no solution for, say, 
  $\bra\Psi_{(h;1)}(z)\Psi_{(h;0)}(w)\ket$
  can be found that satisfies all three Ward identities (\ref{eq:ward}).
  On the other hand, two-point functions are insensitive to non-quasi-primary
  contributions, since these always would result in a two-point function
  of fields of differing conformal weights. Similar inconsistencies appear
  when other cases violating (\ref{jlevel}) are considered. 

  The higher rank case is more complicated, since OPEs might yield terms
  of too high J-level $k$, which do not influence the two-point functions
  as long as $k<r-1$. One has to consider three-point functions:
  Assume that e.g.\ the OPE $\Psi_{(h_1;0)}\Psi_{(h_2;0)}$ produces
  a term of {\em maximal\/} J-level $k$ with conformal weight $h_3$.
  Consider $\bra\Psi_{(h_1;0)}\Psi_{(h_2,0)}\Psi_{(h_3,l)}\ket$
  with $l$ being the {\em minimal\/} J-level such
  that this three-point function is non-zero, and thus has the
  canonical form $C_{(h_1;0)(h_2;0)(h_3;l)}\Pi(z_{ij})^{h_k-h_i-h_j}$.
  Then, one easily finds solutions for 
  $\bra\Psi_{(h_1;0)}\Psi_{(h_2,0)}\Psi_{(h_3,l+1)}\ket$ and
  $\bra\Psi_{(h_1;0)}\Psi_{(h_2,1)}\Psi_{(h_3,l)}\ket$. With these
  solutions as inhomogeneities in the Ward identities (\ref{eq:ward}),
  no consistent solution for 
  $\bra\Psi_{(h_1;0)}\Psi_{(h_2,1)}\Psi_{(h_3,l+1)}\ket$ can be found.
  Thus, we believe that LCFTs without a gradation in the J-level
  are not consistent, although we do not have a complete proof for this
  conjecture. 

\section{\reseteqn{\sc Conclusion}}

  {\sc Taking into account} the proper action of the Virasoro algebra on
  logarithmic fields, i.e.\ working with Jordan cell representations as
  generalizations of irreducible highest-weight representations
  \cite{Rohsiepe:1996}, allows
  to evaluate OPEs in LCFT in a similar fashion as in
  ordinary CFT. The main difference is that each $n$-point function represents
  a full hierarchy of conformal blocks involving $s=1,\ldots,n$ logarithmic
  fields with varying J-levels $k_i=0,\ldots r-1$. While in ordinary conformal
  field theory it suffices to know correlation functions of primary fields only
  (since everything else is fixed by conformal covariance), in logarithmic CFT 
  one needs to know the full hierarchy of $r^n$ correlations functions of 
  primaries and all their logarithmic partner fields. Actually, there are only
  $r^n-{n+r-2\choose r-2}$ different such correlation functions, since the total
  J-level must be at least $r-1$. If one additionally takes into account, that
  all correlation functions with total J-level precisely equal to the minimal
  value $r-1$ are identical (and do not involve any logarithms), the total 
  number of different correlation functions reduces finally to 
  $r^n-{n+r-1\choose r-1} + 1$.

  The solution of this hierarchy can be obtained
  step by step, where the case with one logarithmic field only of maximal
  J-level is worked out in the same way as in ordinary CFT. The same holds
  for correlation functions with several logarithmic fields, such that the
  total J-level adds up to $r-1$. In each further
  step, the differential equations, which result from the existence of
  null vectors, are inhomogeneous, with the inhomogeneity
  determined by the conformal blocks of correlators with fewer logarithmic
  fields. Details on the computation of four-point functions have been
  presented in the last reference in \cite{Flohr:1996a}. 
  Of course, since the OPEs are entirely determined in terms of the two- and 
  three-point functions, the situation is simpler. As we have shown, these 
  are -- completely analogous to the case of ordinary CFT --
  already fixed up to constants by global conformal covariance.
  The formulae (\ref{eq:2pt}) and (\ref{eq:3pt}) yield the full
  hierarchy of these functions in a direct way.

  With the help of a full set of
  two- and three-point functions, OPEs are then simply obtained through
  certain matrix products. Their non-trivial structure is essentially
  due to the fact that the matrix of two-point functions must be
  inverted first. The long example given in section IV demonstrates that
  this inversion results in many additional terms and non-trivial
  linear combinations of the involved structure constants which were
  not accounted for in older approaches, e.g.\ the one taken in 
  \cite{MoghimiAraghi:2001a}.
  In particular, it is not yet clear, how the otherwise elegant method of
  \cite{MoghimiAraghi:2001a}
  to write correlation functions as formal power series expansions in 
  nilpotent conformal weights $h+\eta$, $\eta^r=0$, can be transferred to 
  operator product expansions. Nonetheless, we believe that 
  our discussion on the structure matrix coefficients in terms 
  of polynomials might be of help here.

  This fills one of
  the few remaining gaps to put LCFT on equal footing with better known
  ordinary CFTs such as minimal models. The success of conformal field theory
  is mainly rooted in the fact that correlation functions can be computed 
  effectively and exactly. The basic tools to achieve this are operator 
  product expansions and differential equations due to the existence of 
  null-vectors. Now, all these tools are also available in the logarithmic 
  case. It is worth mentioning the following difference between the
  ordinary and the logarithmic case: In the ordinary case, any OPE 
  $\Phi_{h_i}\Phi_{h_j}$ one wishes to compute depends only on two 
  constants per primary field $\Phi_{h_k}$ occurring on the right hand side, 
  namely $C_{h_ih_jh_k}$ and $D^{h_kh_k}$. This remains true even in the
  case of multiplets of fields of equal conformal weight, as long as their
  two-point functions can be diagonalized. If this latter situation
  does no longer hold, which precisely constitutes the logarithmic CFT
  case \cite{Gurarie:1993}, then things get more complicated. 
  Now, when computing an OPE
  $\Psi_{(h_i;k_i)}\Psi_{(h_j;k_j)}$, one needs for each field 
  from a Jordan block $\{\Psi_{(h_l;k)}\}_{k=0,\ldots,r-1}$ occurring 
  on the right hand side complete knowledge
  of all structure constants $C_{(h_i,h_j,h_l;r-1+s)}$ and
  $D_{(h_l;r-1+t)}$ for $0\leq s\leq k_i+k_j$, $0\leq t\leq{\rm min}(
  r-1,k_i+k_j)$. Thus, for each triplet
  of conformal weights $(h_i,h_j,h_l)$, one needs
  in fact $2r-1$ three-point structure constants plus $r$ two-point
  structure constants, i.e.\ in total $3r-1$ constants, for a rank $r$ LCFT.

  Moreover, we discussed how
  (para-)fermionic and twist fields can be incorporated into our
  framework. A LCFT can be viewed as an extended ordinary CFT where the space
  of states has additional sectors. In particular, we showed that any rank two
  LCFT can consistently accommodate two additional fermionic sectors, and we 
  expect that rank $r$ LCFTs will allow for $\mathbb{Z}_r$ para-fermionic 
  sectors.  Furthermore, LCFTs naturally incorporate twist fields which 
  considerably enrich the structure of such theories. These twist fields 
  can be considered as the basic entities from which all other fields can be 
  generated by successive application of operator products. Since the OPE of 
  a twist with its anti-twist produces, among others, a logarithmic field, 
  twist fields are also called pre-logarithmic fields \cite{Kogan:1998a}. 
  We expect that a more detailed and
  careful treatment of twist fields in higher rank LCFTs will shed new light
  on the still mysterious geometric aspects underlying logarithmic conformal
  field theories (see \cite{Flohr:1998b} for some initial remarks on this).

  We also presented an BRST-like algebraic argument on the zero mode 
  content, which yields necessary conditions for correlators to be non-zero.
  This could be used to consider Jordan cells with non-proper primary
  fields and non-quasi-primary logarithmic partners in more detail.
  In particular, our argument provides conditions for both, the minimal 
  zero mode content of correlators, as well as the maximal zero mode
  content of the right hand side of OPEs, thus yielding a starting point 
  for a generalized hierarchical scheme.

  To conclude, one should remark that we have not yet discussed the most general
  case. Within this paper, we assumed that the only non-vanishing one-point
  function is given by the maximal J-level logarithmic partner of the
  identity, $\bra \Psi_{(h=0;r-1)}\ket$. However, this might be too
  restrictive. In particular, a full discussion of logarithmic CFT should
  include the case of boundaries with their induced spectrum of boundary
  operators, which possess non-vanishing one-point functions.
  It would be
  interesting to compute boundary-boundary OPEs and boundary-bulk OPEs
  in the LCFT case along the lines of our approach.
 
  \bigskip\noindent{{\sc Acknowledgment:}} We thank 
  Shahin Rouhani and in particular Matthias Gaberdiel 
  for many valuable discussions and comments.


\begin{thebibliography}{}
\parsep 0pt
\itemsep 0pt

\bibitem{Bernard:1997}
D.\ Bernard, Z.\ Maassarani, P.\ Mathieu,
\mpla{12}{1997}{535}
[\hepth{9612217}].

\bibitem{Bhaseen:2001}
M.J.\ Bhaseen,
\npb{604}{2001}{537}
[\condmat{0011229}].

\bibitem{Bhaseen:2000a}
M.J.\ Bhaseen, J.-S.\ Caux, I.I.\ Kogan, A.M.\ Tsvelik,
{\it Disordered Dirac Fermions: the Marriage of Three Different Approaches},
[\condmat{0012240}].
\bibitem{Bhaseen:2000b}
M.J.\ Bhaseen,
{\it Random Dirac fermions: The $su(N)$ gauge potential and $\mathbb{Z}_N$ 
     twists},
[\condmat{0012420}]

\bibitem{Bilal:1994a}
A.\ Bilal, I.I.\ Kogan,
\npb{449}{1995}{569}
[\hepth{9503209}].

\bibitem{CampbellSmith:2000a}
A.\ Campbell-Smith, N.E.\ Mavromatos,
\plb{476}{2000}{149}
[\hepth{9908139}];
\plb{488}{2000}{199}
[\hepth{0003262}].

\bibitem{Cappelli:1998}
A.\ Cappelli, L.S.\ Georgiev, I.T.\ Todorov,
\cmp{205}{1999}{657}
[\hepth{9810105}].

\bibitem{Cardy:1999}
J.\ Cardy,
{\it Logarithmic Correlations in Quenched Random Magnets and Polymers},
[\condmat{9911024}].

\bibitem{Caux:1996}
J.-S.\ Caux, I.I.\ Kogan, A.M.\ Tsvelik,
\npb{466}{1996}{444}
[\hepth{9511134}].
\bibitem{Caux:1998a}
J.-S.\ Caux, N.\ Taniguchi, A.M.\ Tsvelik,
\npb{525}{1998}{671}
[\condmat{9801055}].

\bibitem{Caux:1997}
J.-S.\ Caux, I.I.\ Kogan, A.\ Lewis, A.M.\ Tsvelik,
\npb{489}{1997}{469}
[\hepth{9606138}].

\bibitem{Caux:1998b}
J.-S.\ Caux,
\prl{81}{1998}{4196},
[\condmat{9804133}].

\bibitem{Eholzer:1997}
W.\ Eholzer, L.\ Feher, A.\ Honecker
\npb{518}{1998}{669}
[\hepth{9708160}].

\bibitem{Ellis:1996}
J.R.\ Ellis, N.E.\ Mavromatos, D.V.\ Nanopoulos,
\ijmpa{12}{1997}{2639}
[\hepth{9605046}];
\ijmpa{13}{1998}{1059}
[\hepth{9609238}];
\grg{32}{2000}{943}
[\grqc{9810086}]; 
\grg{32}{2000}{1777}
[\grqc{9911055}];
\prd{62}{2000}{084019}
[\grqc{0006004}].
\bibitem{Ellis:1999}
J.R.\ Ellis, N.E.\ Mavromatos, E.\ Winstanley,
\plb{476}{2000}{165}
[\hepth{9909068}].

\bibitem{Flohr:1996a}
M.\ Flohr,
\ijmpa{11}{1996}{4147}
[\hepth{9509166}];
\ijmpa{12}{1997}{1943}
[\hepth{9605151}];
\npb{514}{1998}{523}
[\hepth{9707090}];
{\it Null vectors in logarithmic conformal field theory},
[\hepth{0009137}].

\bibitem{Flohr:1996b}
M.\ Flohr,
\mpla{11}{1996}{55}
[\hepth{9605152}].

\bibitem{Flohr:1996c}
M.\ Flohr,
\npb{482}{1996}{567}
[\hepth{9606130}].

\bibitem{Flohr:1998b}
M.\ Flohr,
\plb{444}{1998}{179}
[\hepth{9808169}].

\bibitem{marco} 
M.\ Flohr, M.\ Krohn, 
in preparation.

\bibitem{Gaberdiel:1996a}
M.R.\ Gaberdiel, H.G.\ Kausch,
\npb{477}{1996}{293}
[\hepth{9604026}];
\plb{386}{1996}{131}
[\hepth{9606050}];
\npb{538}{1999}{631}
[\hepth{9807091}].

\bibitem{Gaberdiel:2001}
M.R.\ Gaberdiel,
{\it Fusion rules and logarithmic representations of a WZW model at  
     fractional level},
[\hepth{0105046}].

\bibitem{Ghezelbash:1997}
A.M.\ Ghezelbash, V.\ Karimipour,
\plb{402}{1997}{282}
[\hepth{9704082}].

\bibitem{Ghezelbash:1999}
A.M.\ Ghezelbash, M.\ Khorrami, A.\ Aghamohammadi,
\ijmpa{14}{1999}{2581}
[\hepth{9807034}].

\bibitem{Giribet:2001}
G.\ Giribet,
\mpla{16}{2001}{821}
[\hepth{0105248}].

\bibitem{Gravanis:2001}
E.\ Gravanis, N.E.\ Mavromatos,
{\it Impulse action on D-particles in Robertson-Walker space times,  
     higher-order logarithmic conformal algebras and cosmological horizons},
[\hepth{0106146}].

\bibitem{Gurarie:1993}
V.\ Gurarie,
\npb{410}{1993}{535}
[\hepth{9303160}].

\bibitem{Gurarie:1997}
V.\ Gurarie, M.\ Flohr, C.\ Nayak,
\npb{498}{1997}{513}
[\condmat{9701212}].

\bibitem{Gurarie:1998}
V.\ Gurarie,
\npb{546}{1999}{765},
[\condmat{9808063}].
\bibitem{Gurarie:1999}
V.\ Gurarie, A.\ Ludwig,
{\it Conformal Algebras of 2D Disordered Systems},
[\condmat{9911392}].

\bibitem{Ino:1997}
K.\ Ino,
\prl{81}{1998}{1078}
[\condmat{980333}];
\prl{82}{1999}{4902}
[\condmat{9812053}];
\prl{83}{1999}{3526}, Erratum \ibid{84}{2000}{201}
[\condmat{9906147}];
\prl{86}{2001}{882}
[\condmat{0008228}].

\bibitem{Ishimoto:2001}
Y.\ Ishimoto,
{\it Boundary states in boundary logarithmic CFT},
[\hepth{0103064}].

\bibitem{Ivashkevich:1998}
E.V.\ Ivashkevich,
\jpha{32}{1999}{1691}
[\condmat{9801183}].

\bibitem{Kausch:1995}
H.G.\ Kausch,
{\it Curiosities at $c$=$-2$},
[\hepth{9510149}];
\npb{583}{2000}{513}
[\hepth{0003029}].

\bibitem{Kaviani:1999}
K.\ Kaviani, A.M.\ Ghezelbash,
\plb{469}{1999}{81}
[\hepth{9902104}].

\bibitem{Kawai:2001}
S.\ Kawai, J.F.\ Wheater,
\plb{508}{2001}{203}
[\hepth{0103197}].

\bibitem{Kheirandish:2001a}
F.\ Kheirandish, M.\ Khorrami,
\epjc{18}{2001}{795}
[\hepth{0007013}];
\epjc{20}{2001}{593}
[\hepth{0007073}].

\bibitem{Khorrami:1998a}
M.\ Khorrami, A.\ Aghamohammadi, M.R.\ Rahimi Tabar,
\plb{419}{1998}{179}
[\hepth{9711155}].

\bibitem{Khorrami:1998b}
M.\ Khorrami, A.\ Aghamohammadi, A.M.\ Ghezelbash,
\plb{439}{1998}{283}
[\hepth{9803071}].

\bibitem{Kim:1998}
J.Y.\ Kim, H.W.\ Lee, Y.S.\ Myung,
{\it Origin of logarithmic corrections in three-dimensional anti-de Sitter 
     space},
[\hepth{9812016}].

\bibitem{Knizhnik:1987}
V.G.\ Knizhnik,
\cmp{112}{1987}{567}.

\bibitem{Kogan:1996a}
I.I.\ Kogan, N.E.\ Mavromatos,
\plb{375}{1996}{111}
[\hepth{9512210}].

\bibitem{Kogan:1996b}
I.I.\ Kogan, C.\ Mudry, A.M.\ Tsvelik,
\prl{77}{1996}{707},
[\condmat{9602163}].

\bibitem{Kogan:1996c}
I.I.\ Kogan, N.E.\ Mavromatos, J.F.\ Wheater,
\plb{387}{1996}{483}
[\hepth{9606102}].

\bibitem{Kogan:1997}
I.I.\ Kogan, A.\ Lewis, O.A.\ Soloviev,
\ijmpa{13}{1998}{1345}
[\hepth{9703028}].

\bibitem{Kogan:1998a}
I.I.\ Kogan, A.\ Lewis,
\npb{509}{1998}{687}
[\hepth{9705240}];
\plb{431}{1998}{77}
[\hepth{9802102}].

\bibitem{Kogan:1999}
I.I.\ Kogan,
\plb{458}{1999}{66}
[\hepth{9903162}].
\bibitem{Kogan:2000c}
I.I.\ Kogan, D.\ Polyakov,
\ijmpa{16}{2001}{2559}
[\hepth{0012128}].

\bibitem{Kogan:2000a}
I.I.\ Kogan, A.M.\ Tsvelik,
\mpla{15}{2000}{931}
[\hepth{9912143}].

\bibitem{Kogan:2000b}
I.I.\ Kogan, J.F.\ Wheater,
\plb{486}{2000}{353}
[\hepth{0003184}].

\bibitem{Kogan:2001b}
I.I.\ Kogan, A.\ Nichols,
{\it $SU(2)_0$ and $OSp(2|2)_{-2}$ WZNW models: Two current algebras, 
     one Logarithmic CFT},
[\hepth{0107160}].

\bibitem{Leontaris:1999}
G.K.\ Leontaris, N.E.\ Mavromatos,
\prd{61}{2000}{124004}
[\hepth{9912230}]; 
\prd{64}{2001}{024008}
[\hepth{0011102}].

\bibitem{Lewis:2000a}
A.\ Lewis,
\plb{480}{2000}{348}
[\hepth{9911163}].

\bibitem{Lewis:2000b}
A.\ Lewis,
{\it Logarithmic CFT on the boundary and the world-sheet},
[\hepth{0009096}].

\bibitem{Ludwig:2000}
A.W.W.\ Ludwig,
{\it A Free Field Representation of the $OSP(2|2)$ Current Algebra at Level 
     $k$=$-2$, and Dirac Fermions in a Random $SU(2)$ Gauge Potential},
[\condmat{0012189}].

\bibitem{Maassarani:1997}
Z.\ Maassarani, D.\ Serban,
\npb{489}{1997}{603}
[\hepth{9605062}].

\bibitem{Mahieu:2001}
S.\ Mahieu and P.\ Ruelle,
{\it Scaling fields in the two-dimensional abelian sandpile model},
[\hepth{0107150}].

\bibitem{Mavromatos:1998}
N.E.\ Mavromatos, R.J.\ Szabo,
\plb{430}{1998}{94}
[\hepth{9803092}].

\bibitem{Mavromatos:1999a}
N.E.\ Mavromatos, R.J.\ Szabo,
\prd{59}{1999}{104018}
[\hepth{9808124}];
{\it D-brane dynamics and logarithmic superconformal algebras},
[\hepth{0106259}].

\bibitem{MoghimiAraghi:2000a}
S.\ Moghimi-Araghi, S.\ Rouhani,
\lmp{53}{2000}{49}
[\hepth{0002142}].

\bibitem{MoghimiAraghi:2001a}
S.\ Moghimi-Araghi, S.\ Rouhani, M.\ Saadat,
\npb{599}{2001}{531}
[\hepth{0008165}];
{\it Current algebra associated with logarithmic conformal field theories},
[\hepth{0012149}].

\bibitem{MoghimiAraghi:2001b}
S.\ Moghimi-Araghi, S.\ Rouhani, M.\ Saadat,
{\it On the AdS/CFT correspondence and logarithmic operator},
[\hepth{0105123}].

\bibitem{Myung:1999}
Y.S.\ Myung, H.W.\ Lee,
\jhep{9910}{1999}{009}
[\hepth{9904056}].

\bibitem{Nichols:2001a}
A.\ Nichols, S.\ Sanjay,
\npb{597}{2001}{633}
[\hepth{0007007}].
\bibitem{Nichols:2001b}
A.\ Nichols,
{\it Logarithmic currents in the $SU(2)_0$ WZNW model},
[\hepth{0102156}].

\bibitem{RahimiTabar:1997a}
M.R.\ Rahimi-Tabar, S.\ Rouhani,
\epl{37}{1997}{447}
[\hepth{9606143}];
{\it Logarithmic Correlation Functions in Two Dimensional Turbulence},
[\hepth{9606154}].

\bibitem{RahimiTabar:1997b}
M.R.\ Rahimi-Tabar, A.\ Aghamohammadi, M.\ Khorrami,
\npb{497}{1997}{555}
[\hepth{9610168}].

\bibitem{RahimiTabar:1998a}
M.R.\ Rahimi-Tabar, S.\ Rouhani,
\plb{431}{1998}{85}
[\hepth{9707060}].

\bibitem{RahimiTabar:1998b}
M.\ Reza Rahimi-Tabar,
\npb{588}{2000}{630}
[\condmat{0002309}]. 

\bibitem{Read:1999}
N.\ Read, D.\ Green,
{\it Paired States of Fermions in Two-Dimensions with Breaking of Parity
and Time Reversal Symmetries, and the Fractional Quantum Hall Effect},
[\condmat{9906453}].

\bibitem{Read:2001}
N.\ Read, H.\ Saleur,
{\it Exact spectra of conformal supersymmetric nonlinear sigma models in two
dimensions},
[\hepth{0106124}].

\bibitem{Rohsiepe:1996}
F.\ Rohsiepe,
{\it On reducible but indecomposable representations of the Virasoro algebra},
[\hepth{9611160}].

\bibitem{Rozansky:1992}
L.\ Rozansky, H.\ Saleur,
\npb{376}{1992}{461};
\npb{389}{1993}{365},
[\hepth{9203069}].

\bibitem{Saleur:1991}
H.\ Saleur,
\npb{489}{1992}{486},
[\hepth{9111007}].

\bibitem{Sanjay:2000}
S.\ Sanjay,
{\it Logarithmic operators in $SL(2,R)$ WZNW model, singletons and  
     $AdS_3/(L)CFT_2$ correspondence},
[\hepth{0011056}].

\bibitem{Skoulakis:1998}
S.\ Skoulakis, S.\ Thomas,
\plb{438}{1998}{301}
[\condmat{9802040}].

\end{thebibliography}
\end{document}